\documentclass[useAMS,usenatbib]{mn2e}
\usepackage{graphicx}
\usepackage{longtable,lscape}
\usepackage{amssymb}
\usepackage[usenames]{color}

\def\aap{A\&A}

\def\apss{Ap\&SS}
\def\apj{ApJ}
\def\apjs{ApJS}
\def\mnras{MNRAS}

\title[Study of sdO models. Pulsation Analysis]{Study of sdO models. Pulsation Analysis}
\author[C. Rodr\'\i guez-L\'opez, A. Moya, R. Garrido, J. MacDonald, R. Oreiro and A. Ulla]{C. Rodr\'\i guez-L\'opez$^{1,2,3}$\thanks{E-mail:crodrigu@ast.obs-mip.fr} A. Moya$^{3}$, R. Garrido$^{3}$, J. MacDonald$^{4}$, R. Oreiro$^{5}$ and A. Ulla$^{2}$ \\
$^{1}$Laboratoire d'Astrophysique de Toulouse-Tarbes, Universit\'e de Toulouse, CNRS, Toulouse 31400, France \\
$^{2}$Departamento de F\'\i sica Aplicada, Universidade de Vigo, Vigo 36310, Spain \\
$^{3}$Departamento de F\'isica Estelar, Instituto de Astrof\'\i sica de Andaluc\'\i a-CSIC, Granada 18008, Spain\\
$^{4}$University of Delaware, Department of Physics and Astronomy, Newark, DE 19716, USA \\
$^{5}$Institute of Astronomy, Katholieke Universiteit Leuven, Celestijnenlaan 200D, 3001 Leuven, Belgium} 

\begin{document}

\date{Accepted 2009 Month dd. Received 2009 Month dd; in original form 2009 Month dd}

\pagerange{\pageref{firstpage}--\pageref{lastpage}} \pubyear{2009}

\maketitle

\label{firstpage}

\begin{abstract}

We have explored the possibility of driving pulsation modes in models of sdO stars in which the effects of element diffusion, gravitational settling and radiative levitation have been neglected so that the distribution of iron-peak elements remains uniform throughout the evolution. The stability of these models was determined using a non-adiabatic oscillations code. We analysed 27 sdO models from 16 different evolutionary sequences and discovered the first ever sdO models capable of driving high-radial order {\it g}-modes. In one model, the driving is by a classical $\kappa$-mechanism due to the opacity bump from iron-peak elements at temperature $\sim 200,000$~K. In a second model, the driving result from the combined action of $\kappa$-mechanisms operating in three distinct regions of the star: (i) a carbon-oxygen partial ionization zone at temperature $\sim 2~10^6$~K, (ii) a deeper region at temperature $\sim 2~10^7$~K, which we attribute to ionization of argon, and (iii) at the transition from radiative to conductive opacity in the core of the star.

\end{abstract}

\begin{keywords}
stars: oscillations -- stars: variables: other
\end{keywords}

\section{Introduction}

Hot subluminous O-stars (hereafter sdOs) are evolved objects on the way to becoming low-mass white dwarfs. Most sdOs are thought to have an inert carbon-oxygen core, with the luminosity provided by helium and hydrogen burning shells. The wide ranges of surface parameters, from 40 to 100~kK for effective temperature ($T_\mathrm{eff}$) and 4.0 to 6.5 for surface gravity ($\log g$), spreads sdOs across a broad region in the HR diagram, that is crossed by post Asymptotic Giant Branch (post-AGB), post Red Giant Branch (post-RGB) and post Extended Horizontal Branch (post EHB) single star evolutionary tracks. This degeneracy in evolutionary paths plus possible binary star formation scenarios hampers reliable determination of sdO origins. This situation is not helped by the wide variety in sdO spectra: While they share a common characteristic of a He~II 4686\AA ~line, a diversity of lines with different strengths of He~I, He~II and sometimes H are possible, rendering  spectral classification an open issue (\citealt{jeffery97}; \citealt{stroer07}).

The spectral analysis and determination of physical parameters has the drawback of requiring the use of non-LTE atmospheric codes due to high effective temperatures, which yielded disagreements in the largest existent studies (\citealt{dreizler90}; \citealt{thejll94}) until recently. Recent analyses of 130 sdOs by \citet{stroer07} and \citet{hirsch08} showed that they segregate into two main groups regarding helium content: (1) {\em helium-deficient} sdOs with subsolar He abundances, no traces of carbon or nitrogen and a large spread in $T_\mathrm{eff}$ and $\log g$, and (2) {\em helium-enriched} sdOs with supersolar He abundances and C and/or N lines present, spreading between $\sim$40-50~kK in $T_\mathrm{eff}$ and 5.5 to 6 in $\log g$. Group (1) sdOs are classified as post-EHB descendants of hot subdwarf B-stars (hereafter sdBs). To explain the origins of group (2) sdOs, the authors invoke different channels: (i) merger of two helium white dwarfs formed by binary evolution (\citealt{webbink84}; \citealt{han02}), which explains the He enrichment, but does not quite reproduce the observed $T_\mathrm{eff}$-$\log g$ distribution, (ii) non-canonical evolution of a single star that experiences a phase of mixing from the H-shell into the envelope while on the RGB \citep{sweigart97} and (iii) evolution of a RGB star whose envelope is almost completely stripped by Roche-lobe overflow in a binary, pre-emptying the He - core flash \citep{driebe99}. This last scenario has problems in explaining the observed metal enrichment. (See \citet{stroer07} and references therein for a thorough review of sdO formation mechanisms).
 
Another way to attack the problem of untangling the different evolutionary scenarios proposed for sdOs, and in which this paper focuses, is the use of asteroseismological methods (provided, of course, that sdOs are pulsating stars). Asteroseismological analysis can yield information such as the mass of the star, envelope, and inner structure through the Brunt-V\"ais\"al\"a frequency and/or sound speed, as long as we can identify the oscillation modes present in the star.

The first theoretical exploration of the potential of sdOs as pulsators was carried out by \citet{charpinet97a}. The stability analysis of full stellar sdO models corresponding to seven evolutionary sequences starting on the EHB found all sdO models stable.

The previous year, \citet{charpinet96} had theoretically predicted the pulsator behaviour of sdBs due to a $\kappa$-mechanism caused by iron enhancement in the envelope of the star. Since the observational confirmation of the first pulsating sdB star by \citet{kilkenny97}, we now know of about forty of these stars \citep{oreiro09} termed V361 Hya, or more commonly EC\,14026 stars, which coexist in the same area of the HR diagram with non-pulsating sdBs. They cluster in $T_\mathrm{eff}$ at $\sim$33\,000 K and $\log g$ $\sim$5.8 dex and show oscillations with typical periods in the range 80--200 s, which have been identified with low-degree and low-radial order {\it p}-modes. Today, the sdB variable stars also include the long period pulsators V1093~Her \citep{green03} also excited by the same $\kappa$ mechanism \citep{fontaine03}, a handful of hybrid pulsators (e.g. \citealt{oreiro05}) and a unique He-sdB star \citep{ahmad05}. For more details on sdB stars we refer the reader to the excellent review by \citet{ostensen09}.

The goal of the present work is to complete the observational search for sdO pulsators (\citealt*{crl07}; \citealt{crl05}; \citealt{crl04}) that proceeded in the period 2003--2006, and which has yielded non-conclusive results. The thrilling discovery of the first sdO pulsator, SDSS\,J160043.6+074802.9, (abbreviated J1600+0748) by a team of South African astronomers \citep{woudt06}, gave real support to our theoretical exercise. The recent discovery of {\em p}-mode oscillations in sdO models that included radiative levitation of iron \citep{fontaine08} is able to account for the the very fast multiperiodic oscillations observed in the range $\sim$60 to $\sim$120~s, hence establishing the same iron opacity mechanism found in sdB stars as responsible for sdO oscillations, and reinforcing the conclusion that radiative levitation and gravitational settling are two key ingredients in driving pulsations (\citealt{charpinet97b}; \citealt*{charpinet09}).

   \begin{figure}
   \includegraphics[width=9cm]{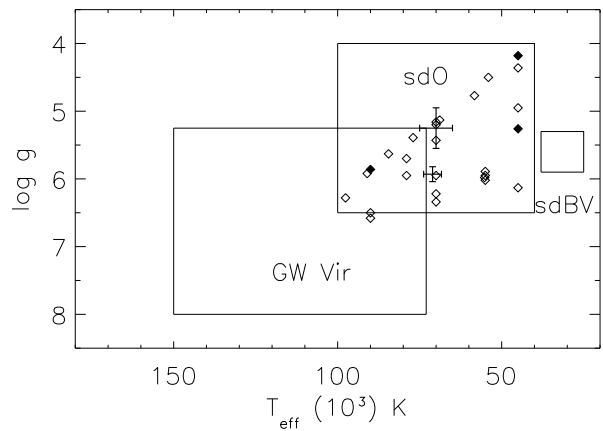}
      \caption{Diamonds mark the location of the sdO models in the $\log g$--$T_\mathrm{eff}$ plane. Filled diamonds correspond to models 8, 10 and 15 which will be thoroughly described in following sections. The error bars give the $T_\mathrm{eff}$ and $\log g$ determination for J\,1600+0748 by \citet{fontaine08} (bottom) and \citet{crl09a} (top). A rough sketch of the sdO domain and the sdBV and GW Vir instability strips is also given.
              }
         \label{fig:models2}
   \end{figure}

Our aim is to investigate the possibility of driving pulsation modes in theoretical models of sdO stars -- coming from fully evolutionary sequences-- with the aid of a non-adiabatic oscillations code. Thus, sdO models were specifically built using the evolutionary code {\scriptsize JMSTAR} (\citealt{lawlor06} and references therein). The code was used to further evolve these sdO models to more advanced state of evolution, providing the tracks taken in the HR diagram. The sdO models are used as input to the {\scriptsize GraCo} non-adiabatic oscillations code (\citealt*{moya04}; \citealt{moya08}), which predicts which oscillation modes of the models are excited. 

The paper is structured as follows: In Section 2 we give some general remarks about the evolution code, the structural models and the non-adiabatic oscillations code. Section 3 briefly describes the non-adiabatic stability analysis carried on the sdO models. Sections 4, 5 and 6 describe the analysis for the most interesting models found. We summarise all the analysed models in Section 7, and the discussion and conclusions in Section 8.

\section{The tools}

\subsection{The evolution code and the sdO models}

To have accurate theoretical models of sdO stars is of great importance in the subsequent pulsation analysis. The distribution of the thermodynamical quantities, such as pressure, temperature, density, and the chemical composition in the interior of the star, will determine the theoretical spectrum of modes computed by the non-adiabatic oscillations code.

\begin{table*}
\centering
\caption{Main physical parameters and mass fractions of the sdO models (models named .1 means they belong to the same evolutionary sequence). X(other) refers to the mass fractions of all the other elements. Z is the current metallicity of the model (in brackets the initial metallicity, when different, at the ZAMS). Models between horizontal lines have been calculated with subtle differences in the code of evolution. (Model names are chosen mainly according to their mass loss rate parameter and effective temperature).}
\label{tab:models}						      
\setlength{\tabcolsep}{3pt}
\begin{tabular}{l|cccccccccccc}  
\hline
Model number &  $T_\mathrm{eff}$ & $\log g$ &  M  & $\eta_\mathrm{R}$ & X(H) & X(He$^3$) &  X(He$^4$) & X(C) &  X(N) & X(O) & X(other) & Z \\ 
\& name      &       (K)         &          & (M$_\odot$)&            &      &  	 &	 &	 &	     &      &	    &	\\ \hline
1  p675\_8057 & 79\,000 & 5.70 & 0.478 & 0.675 & 0.21 & 6.1E-05 & 0.72 & 3.4E-02 & 1.0E-02 & 5.2E-03 & 1.7E-02 & 0.07 (0.02) \\
1.1  p675\_9065 & 90\,000 & 6.50 & 0.478 & 0.675 & 0.20 & 5.4E-05 & 0.73 & 3.5E-02 & 1.0E-02 & 5.1E-03 & 1.7E-02 & 0.07 (0.02) \\
2  p650\_7960 & 79\,000 & 5.95 & 0.491 & 0.650 & 0.68 & 2.4E-04 & 0.30 & 3.0E-03 & 1.7E-03 & 1.0E-02 & 5.3E-03 & 0.02 \\
\hline						        	  	 
3  etap\,685  & 55\,000 & 5.89 & 0.471 & 0.685 & 0.43 & 1.4E-04 & 0.55 & 9.6E-04 & 8.1E-03 & 5.6E-03 & 5.3E-03 & 0.02  \\
4  etap\,690  & 55\,000 & 5.95 & 0.471 & 0.690 & 0.33 & 4.5E-05 & 0.65 & 6.3E-04 & 1.0E-02 & 4.0E-03 & 5.3E-03 & 0.02  \\
5  etap\,695  & 55\,200 & 5.98 & 0.471 & 0.695 & 0.27 & 2.8E-05 & 0.71 & 9.0E-04 & 1.1E-02 & 3.2E-03 & 5.4E-03 & 0.02  \\
6  etap\,700  & 55\,000 & 6.02 & 0.470 & 0.700 & 0.18 & 5.6E-05 & 0.78 & 1.5E-02 & 7.8E-03 & 2.6E-03 & 9.8E-03 & 0.04 (0.02)   \\
\hline							       
7  eta\,600t45 & 45\,000 & 4.95 & 0.497 & 0.600 & 0.59 & 2.5E-04 & 0.36 & 7.3E-03 & 4.2E-03 & 2.4E-02 &1.5E-02 & 0.05  \\
7.1  eta\,600t70 & 70\,000 & 5.95 & 0.497 & 0.600 & 0.59 & 1.9E-04 & 0.35 & 7.3E-03 & 4.3E-03 & 2.4E-02 & 1.5E-02 & 0.05  \\
8  eta\,650t45 & 45\,000 & 5.26 & 0.469 & 0.650 & 0.59 & 1.2E-03 & 0.35 & 7.4E-03 & 4.2E-03 & 2.4E-02 &1.5E-02 & 0.05  \\
8.1 eta\,650t70 & 70\,000 & 6.22 & 0.468 & 0.650 & 0.59 & 1.1E-03 & 0.36 & 7.4E-03 & 4.2E-03 & 2.4E-02 & 1.5E-02 & 0.05 \\
9  eta\,700t45 & 45\,000 & 6.13 & 0.454 & 0.700 & 1.6E-03 & 8.2E-07 & 0.93 & 1.1E-02 & 2.7E-02 & 3.7E-03 & 2.7E-02 &  0.07 (0.05)  \\
9.1 eta\,700t70 & 70\,000 & 6.34 & 0.453 & 0.700 & 1.63E-03 & 3.6E-07 & 0.93 & 1.1E-02 & 2.7E-02 & 3.7E-03 & 2.7E-02 & 0.07 (0.05) \\
\hline							       
10 eta\,675t45mixi   & 45\,000 & 4.18 & 0.479 & 0.675 & 0.26 & 5.2E-05 & 0.60 & 6.0E-02 & 9.5E-03 & 5.4E-02 & 1.7E-02 & 0.14 (0.02)  \\
10.1 15773t54g45 & 54\,500 & 4.54 & 0.479 & 0.675 & 0.24 & 4.3E-05 & 0.61 & 6.3E-02 & 9.7E-03 & 5.6E-02 & 1.7E-02 & 0.15 (0.02) \\ 
10.2 15873t58g47 & 58\,300 & 4.68 & 0.479 & 0.675 & 0.24 & 4.1E-05 & 0.62 & 6.4E-02 & 9.8E-03 & 5.7E-02 & 1.7E-02 & 0.14 (0.02) \\ 
10.3 16973t69g51 & 69\,000 & 5.14 & 0.479 & 0.675 & 0.19 & 2.4E-05 & 0.65 & 7.1E-02 & 1.0E-02 & 6.2E-02 & 1.9E-02 & 0.16 (0.02) \\ 
10.4 17773t77g54 & 77\,000 & 5.39 & 0.479 & 0.675 & 0.16 & 1.4E-05 & 0.67 & 7.7E-02 & 1.1E-02 & 6.7E-02 & 2.0E-02 & 0.17 (0.02) \\ 
10.5 18273t84g56 & 84\,500 & 5.63 & 0.479 & 0.675 & 0.13 & 9.0E-06 & 0.69 & 8.1E-02 & 1.1E-02 & 7.0E-02 & 2.1E-02 & 0.18 (0.02) \\ 
10.6 18573t91g59 & 90\,700 & 5.93 & 0.479 & 0.675 & 0.11 & 5.9E-06 & 0.70 & 8.4E-02 & 1.2E-02 & 7.3E-02 & 2.1E-02 & 0.19 (0.02) \\ 
11 eta\,675t45mixnmi & 45\,000 & 4.36 & 0.480 & 0     & 0.48 & 3.0E-04 & 0.45 & 2.8E-02 & 5.9E-03 & 2.9E-02 & 1.0E-02 & 0.07  (0.02) \\
12 eta\,675t70mix1i  & 70\,000 & 5.16 & 0.479 & 0.675 & 0.19 & 2.3E-05 & 0.65 & 7.1E-02 & 1.1E-02 & 6.2E-02 & 1.9E-02 & 0.16 (0.02)  \\
13 eta\,675t70mix2i  & 70\,000 & 5.20 & 0.479 & 0.675 & 0.16 & 1.6E-05 & 0.66 & 7.5E-02 & 1.1E-02 & 6.6E-02 & 2.0E-02 & 0.18 (0.02)  \\
14 eta\,675t70mixnmi & 70\,000 & 5.43 & 0.480 & 0     & 0.48 & 3.0E-04 & 0.45 & 2.8E-02 & 5.9E-03 & 2.9E-02 & 1.0E-02 & 0.07 (0.02)  \\
15 eta\,675t90mixi   & 90\,000 & 5.86 & 0.479 & 0.675 & 0.11 & 6.2E-06 & 0.70 & 8.4E-02 & 1.2E-02 & 7.3E-02 & 2.1E-02 & 0.19 (0.02)  \\
15.1 18742t98g63 & 97\,600 & 6.28 & 0.479 & 0.675 & 0.09 & 3.4E-06 & 0.71 & 8.8E-02 & 1.2E-02 & 7.6E-02 & 2.2E-02 & 0.20 (0.02)   \\
16 eta\,675t90mixnmi & 90\,000 & 6.58 & 0.480 & 0     & 0.48 & 3.0E-04 & 0.45 & 2.8E-02 & 5.9E-03 & 2.9E-02 & 1.0E-02 & 0.07 (0.02)  \\
\hline
\end{tabular}							       
\centering							       
\end{table*}

The evolution code {\scriptsize JMSTAR}  was used to calculate a total of 16 initial sdO models corresponding to different evolutionary sequences (Table~\ref{tab:models} shows some of their relevant properties). The models were built in a first step to account for the physical parameters of the pulsator sdOs candidates found in our simultaneous observational survey, and in a second step, trying to map out the sdO domain in the HR diagram. All of the models were created by evolving a 1 M$\odot$ star, starting on the pre-Main Sequence. The initial metallicity is solar (Z = 0.02) except models 7--9 for which Z = 0.05. The code follows the complete evolution of the star continuously to the sdO stage, and beyond. Fig.~\ref{fig:models2} roughly depicts the sdO domain in the HR diagram and the location of the constructed models. Their domain is located between the instability strips of the sdB variables and the GW\,Vir stars (low-degree {\em g}-mode pulsators excited by the C/O opacity mechanism)

To obtain sdO models, the mass loss rate on the RGB is enhanced by a factor ranging from 1.60 to 1.85 compared to canonical Reimers' mass loss rates. Most of the hydrogen-rich envelope is removed just after or even before the helium core flash. At the highest RGB rates, the model stars leave the RGB and evolve to higher temperatures at constant luminosity, avoiding the AGB. They then suffer a helium-core flash just before or slightly after they enter a white dwarf cooling phase. After the core flash, the star is an EHB/sdB star burning He quiescently in the core. After core He exhaustion, the star enters the He shell burning sdO phase. The calculations are stopped when the model stars reach the (final) white dwarf phase. A born-again phase due to a very late thermal pulse occurs in some of the sequences. In this case, flash-driven convection carries nuclear processed material to the surface of the star, enriching the atmosphere with He, C, N and O. For particular models with sdO characteristics (see Table~\ref{tab:models}), we have performed a stability analysis.

\begin{figure}
\includegraphics[width=9cm]{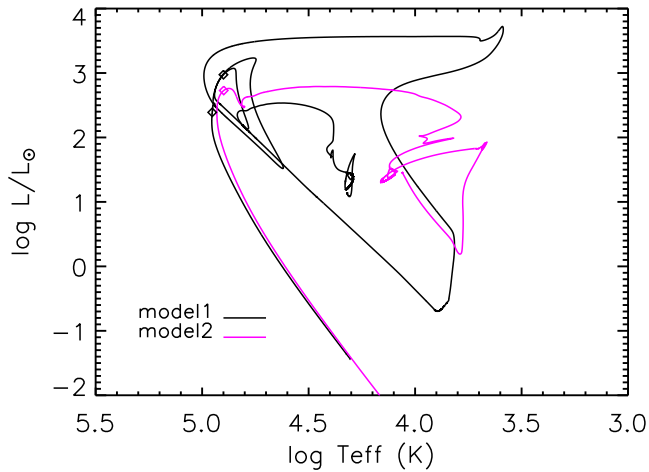} 
\caption{Evolutionary tracks from the horizontal branch described by models 1 and 2. A stability analysis was done also for a subsequent model in the evolutionary track of model 1, namely, model 1.1. Diamonds mark the location of the models.}
\label{fig:dhrp675p650}
\end{figure}

\begin{figure}
\includegraphics[width=9cm]{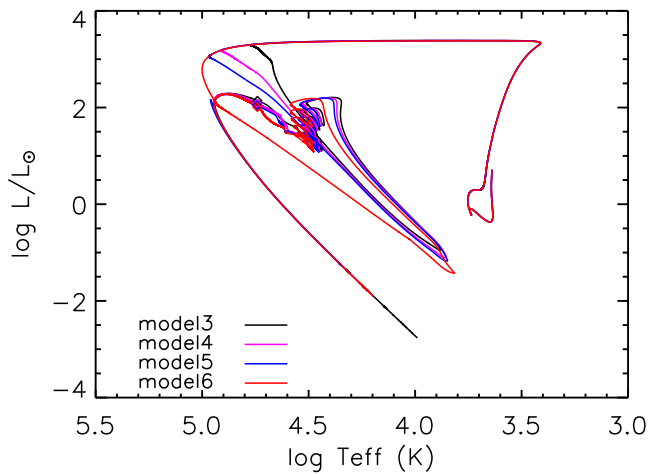} 
\caption{Complete evolutionary tracks for models 3--6. Diamonds mark the location of the models.}
\label{fig:dhr2}
\end{figure}

\begin{figure}
\includegraphics[width=9cm]{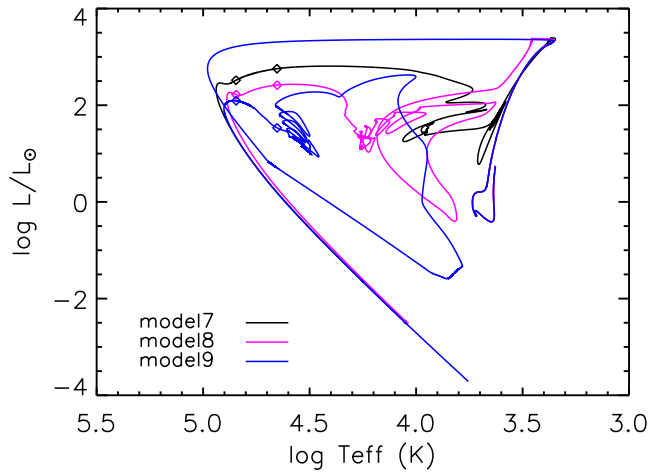} 
\caption{Complete evolutionary tracks for models 7--9. A stability analysis was done also for subsequent models in the evolutionary track, namely, models 7.1, 8.1 and 9.1. Diamonds mark the location of the six models.}
\label{fig:dhrz05}
\end{figure}

\begin{figure}
\includegraphics[width=7.5cm,angle=-90]{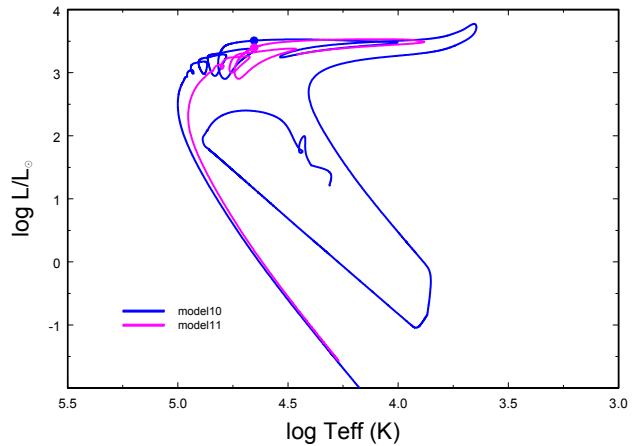} 
\caption{Evolutionary tracks from the horizontal branch for models 10 and 11. Dots mark the location of the models.}
\label{fig:gtextramix}
\end{figure}

 The evolutionary tracks on which models 1--2, 3--6, 7--9 and 10--11 lie are depicted in Fig.~\ref{fig:dhrp675p650},~\ref{fig:dhr2},~\ref{fig:dhrz05} and~\ref{fig:gtextramix}. The complex evolution, near $T_\mathrm{eff}$ = 30\,000~K ($\log T_\mathrm{eff} \sim 4.5$) in models 3 to 6, and in a wider range of temperatures for models 7--9, occurs near the end of the helium-core flash when the star is settling to a quiescent helium burning phase in the core, that is, the sdB phase. At the high mass loss rates used to create models 3--6, the mass loss induced by the hot star wind is sufficient to expose internal He-rich layers. Thus, although the photospheres are not He-rich when the models reach the Horizontal Branch, they become He enriched due to the stellar wind as they evolve from sdBs to sdOs.  Note that Fig.~\ref{fig:dhrp675p650} shows only the part of the evolutionary tracks that occurs after the EHB phase.

\begin{figure}
\includegraphics[width=8.5cm,angle=-90]{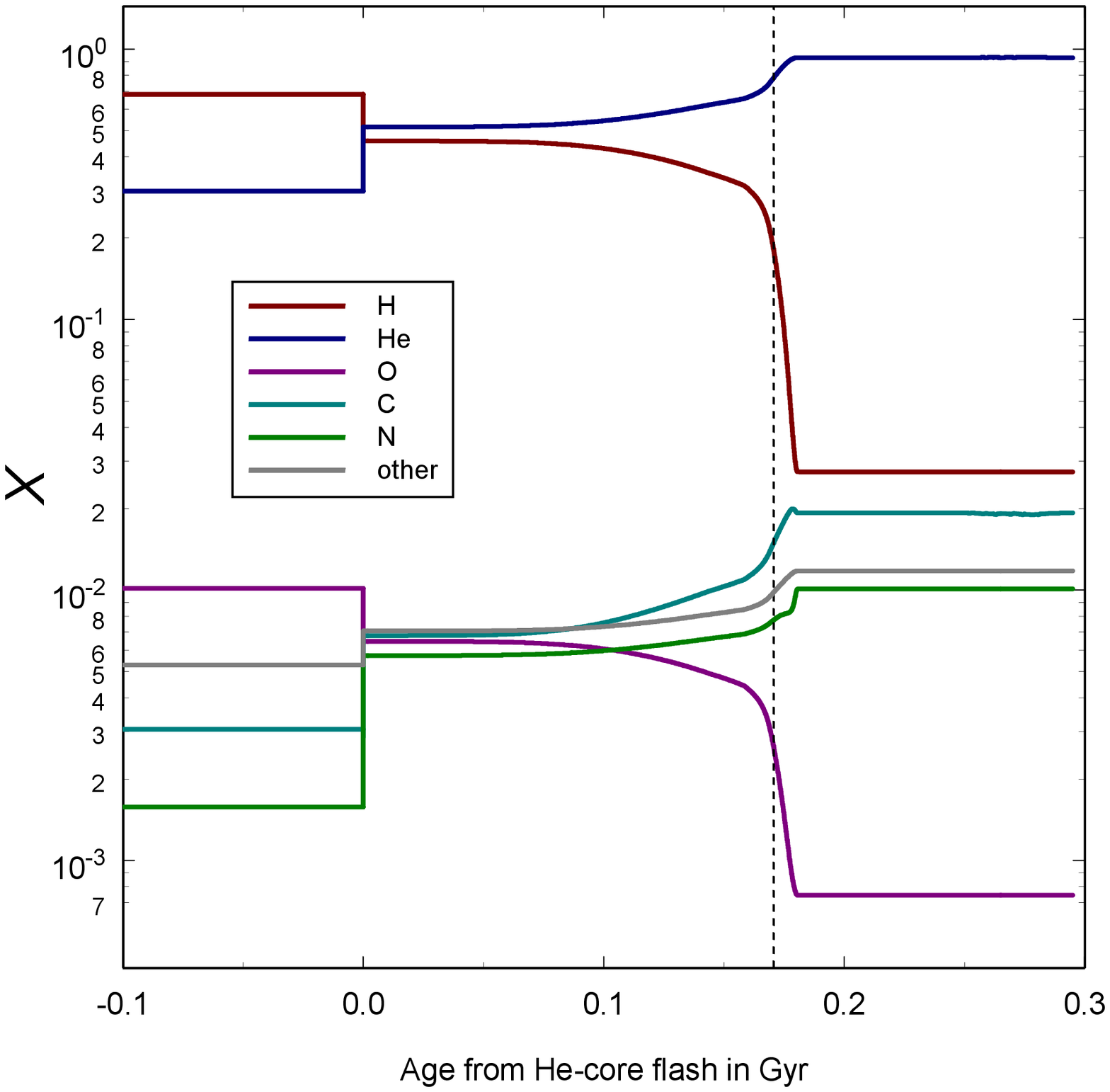}
\caption{Temporal evolution of the photospheric abundances of the model 6 sequence. Time is measured from the peak of the He-core flash. The vertical line indicates the location of the sdO model 6.}
\label{fig:sab6}
\end{figure}

There is an important difference between the evolutionary sequence of model 6 and the similar sequences of models 3--5. For model 6, the He-core flash occurs after the star has entered the white dwarf cooling phase, and in a similar manner to the evolution of post-AGB 'Born-Again' stars (see e.g. \citealt{lawlor06} and references therein), the convection zone that develops above the off-center He-burning region penetrates into the H-rich envelope. However, unlike the 'Born-Again' stars, this occurs when the He-burning region has become inactive and the temperature is too low for proton capture reactions to occur. The result of the mixing is to introduce C$^{12}$ produced by the 3-$\alpha$ process into the envelope, together with O$^{18}$ and Ne$^{22}$ produced by N$^{14}$ $\alpha$-captures. Shortly after the end of the He-core flash, this material is convected throughout the H-rich envelope, increasing the heavy element abundance to $Z = 0.04$. For models 3--5, the convection zone above the off-center He-burning region does not penetrate into H-rich envelope and there is no increase in the surface heavy element abundances. The temporal evolution of the photospheric abundances for the model 6 sequence is shown in Fig.~\ref{fig:sab6}. 

Models 10 to 16 require special comment. They are also late core helium flashers, but after they had arrived on the Zero Age Horizontal Branch (ZAHB) ad hoc extra-mixing was included for the purpose of bringing heavy elements to the surface to increase the envelope metallicity. Due to the extra-mixing, helium is also taken to higher temperatures, leading to helium shell flashes. Flash-driven convection then leads to further mixing of heavy elements to the surface. These helium shell flashes are responsible for the loops in Fig.~\ref{fig:gtextramix}, near $T_\mathrm{eff} \sim 60\,000$~K and $\log g \sim 5$. For models 11 (see Fig.~\ref{fig:gtextramix}), 14 and 16 mass loss was switched off in the calculations once they arrived on the ZAHB.

\subsection{The oscillations code}

We used the {\scriptsize GraCo} fully non-adiabatic oscillations code  (\citealt{moya04}; \citealt{moya08}), which is based on the \citet{unno89} formulation to study the sdO models described above.
Because the thermal and dynamical time scales become comparable in the outer parts of the star, a non-adiabatic treatment of the oscillations is required in order to properly describe the dynamical characteristics of the oscillations, {\em i.e.} which modes will grow or decay with time. The non-adiabatic treatment does not significantly modify the eigenfrequencies with respect to the adiabatic, at least for our purposes, although it will modify the amplitudes and phases of the eigenfunctions. This proves to be very important in using multicolor photometry for mode identification of pulsating stars, although this is beyond the scope of this paper.

\section{Non Adiabatic Stability Analysis}

We used {\scriptsize GraCo} to calculate modes from $l=0$ to $l=4$ with frequencies between $\sim$0.3 to $\sim$20~mHz ($\sim$50 to 3000~s), searching for excited short and/or long periods.

The stability or instability of a model is given by the growth rate as a function of the frequency. We used the \citet*{dziembowski93} definition of the growth rate:
$\eta= W / \int_{0}^{R}|dW/dr|dr$ with the convention that positive (negative) values indicate driving (damping) of a particular mode.

In discussing which regions of the model contribute to driving or damping, we made use of the derivative of the work integral as a function of the fractional mass depth\footnote{The fractional mass depth parameter $\log q = \log (1- M_r/M_T$) gives a better resolution of the envelope structure, which is more relevant to pulsations than the structure of the deep interior.}  ($dW /d \log q$), which gives the energy locally gained or lost by the displaced material during one pulsation cycle. A positive (negative) value of $-dW/d \log q$ at a given location indicates that this region contributes locally to driving (damping) a mode. Correspondingly, the running work integral ($-W$) is also used to illustrate the relative importance of driving and damping zones.

In the next sections we describe the more interesting models found in the analysis.

\section{Model 8: eta\,650t45}

This model, taken from an evolutionary sequence with initial heavy element abundance Z = 0.05, has $T_\mathrm{eff}$=45\,000~K and $\log g=5.25$. Although it was found stable in the whole frequency range, it has a narrow intermediate frequency region, corresponding to low-radial order {\em p}-modes, with a weak tendency to instability, which seemed interesting in terms of explaining a possible selection mechanism for observed modes in the only sdO pulsator found to date (\citet{crl09b}.

   \begin{figure}
   \includegraphics[width=9cm]{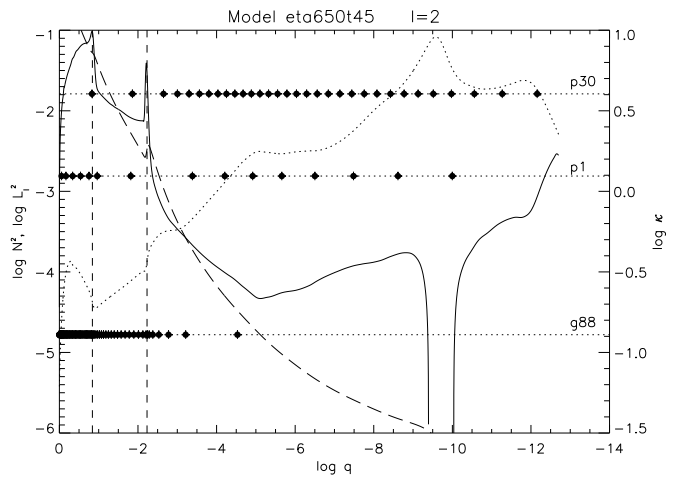} 
   \caption{Propagation diagram: Brunt-V\"ais\"al\"a (solid line) and Lamb frequency (long-dashed line) for $l=2$ modes of model 8. The opacity (dotted line) scale is given in the right axis. The vertical dashed lines mark the maximum of the C/O-He and He-H chemical transitions. The frequencies of a pure {\em g}-mode ({\em g}88), a mixed mode ({\em p}1) and an almost pure {\em p}-mode ({\em p}30) are given, with the positions of their nodes marked by circles. See text for further details.}
   \label{fig:propaga}
   \end{figure}

   \begin{figure*}
   \begin{tabular}{cl}
   \resizebox{0.32\linewidth}{1.5in}{\includegraphics[angle=90]{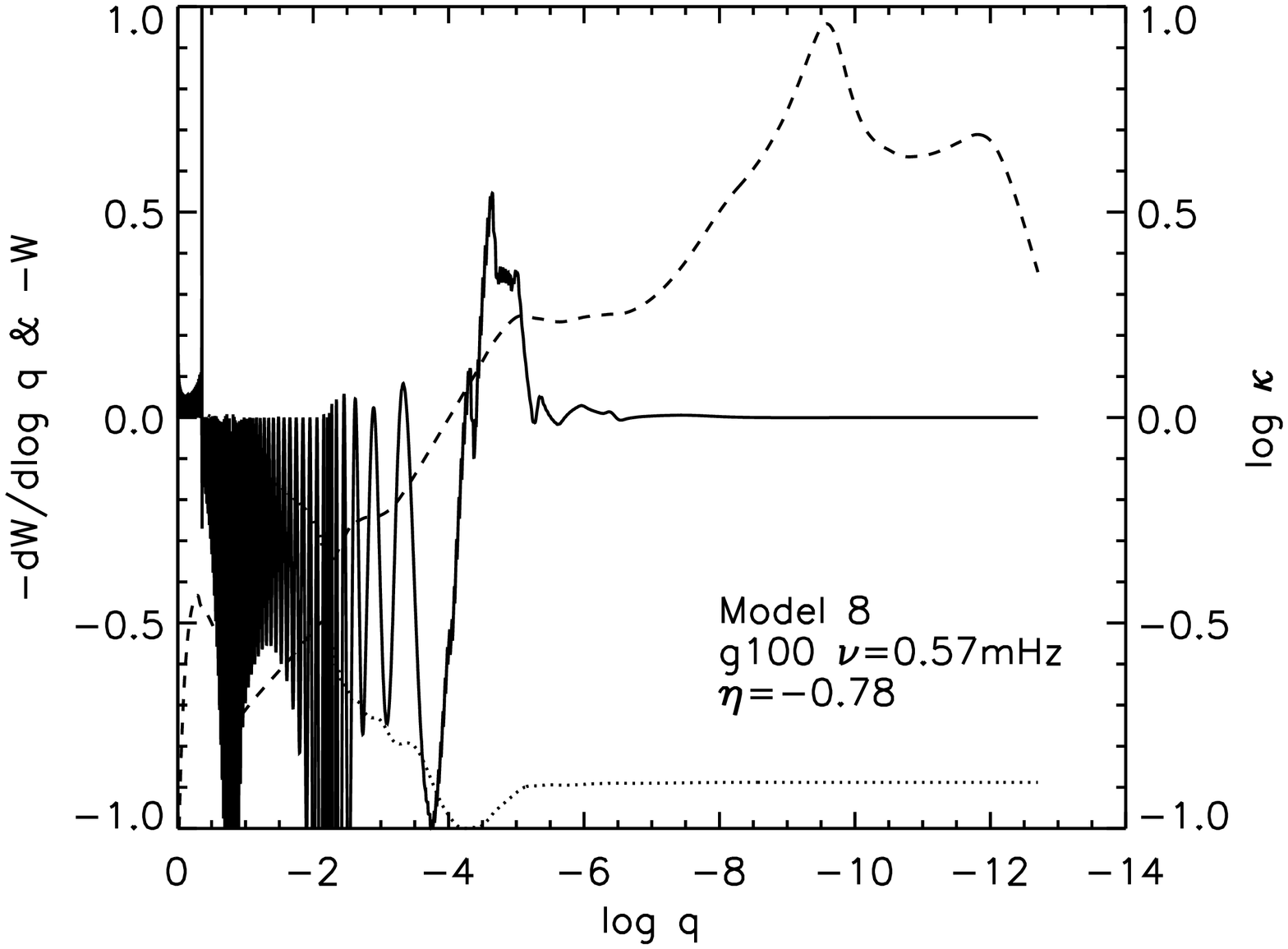}} 
   \resizebox{0.32\linewidth}{1.5in}{\includegraphics[angle=90]{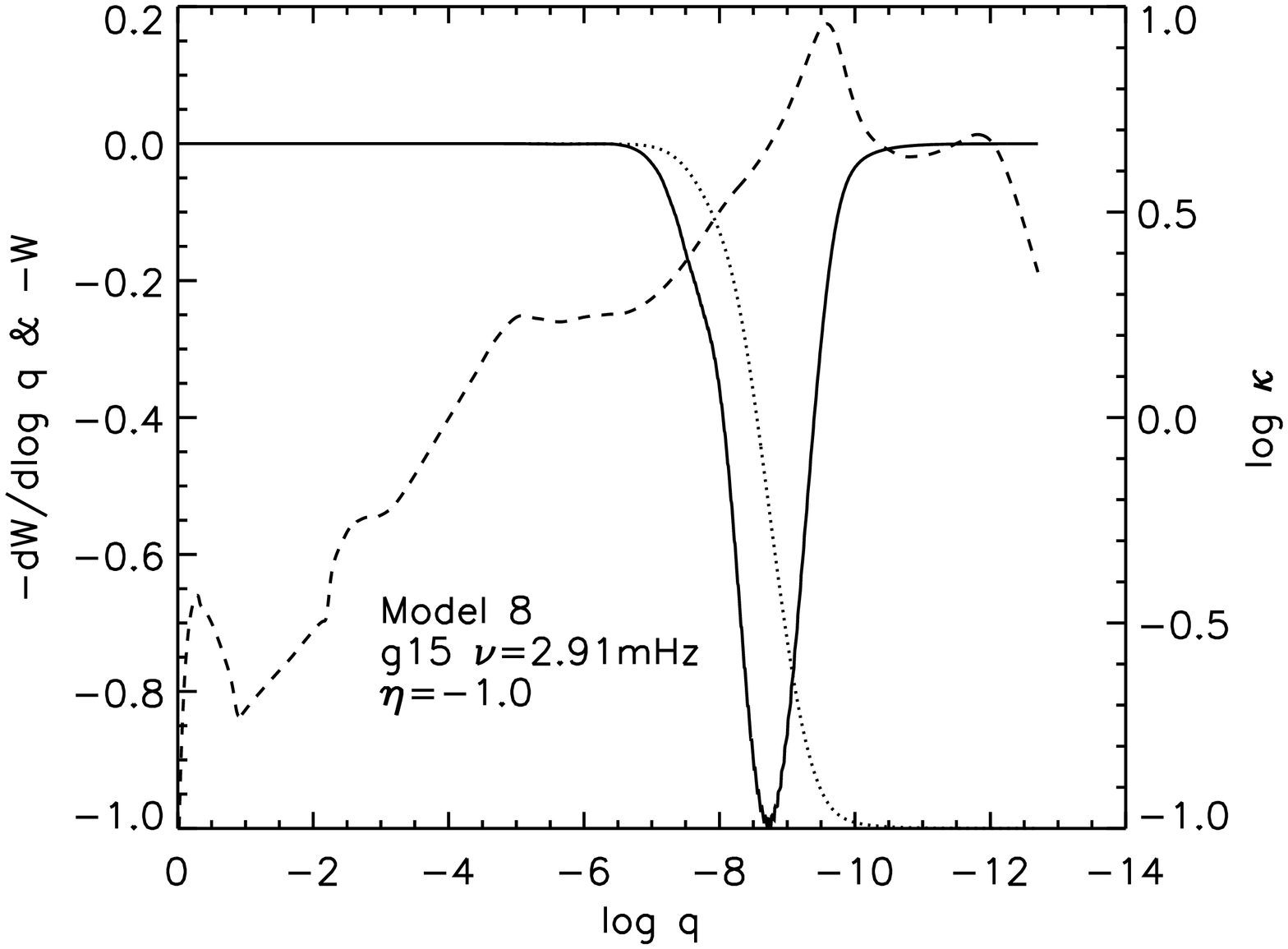}} &
   \resizebox{0.32\linewidth}{1.5in}{\includegraphics[angle=90]{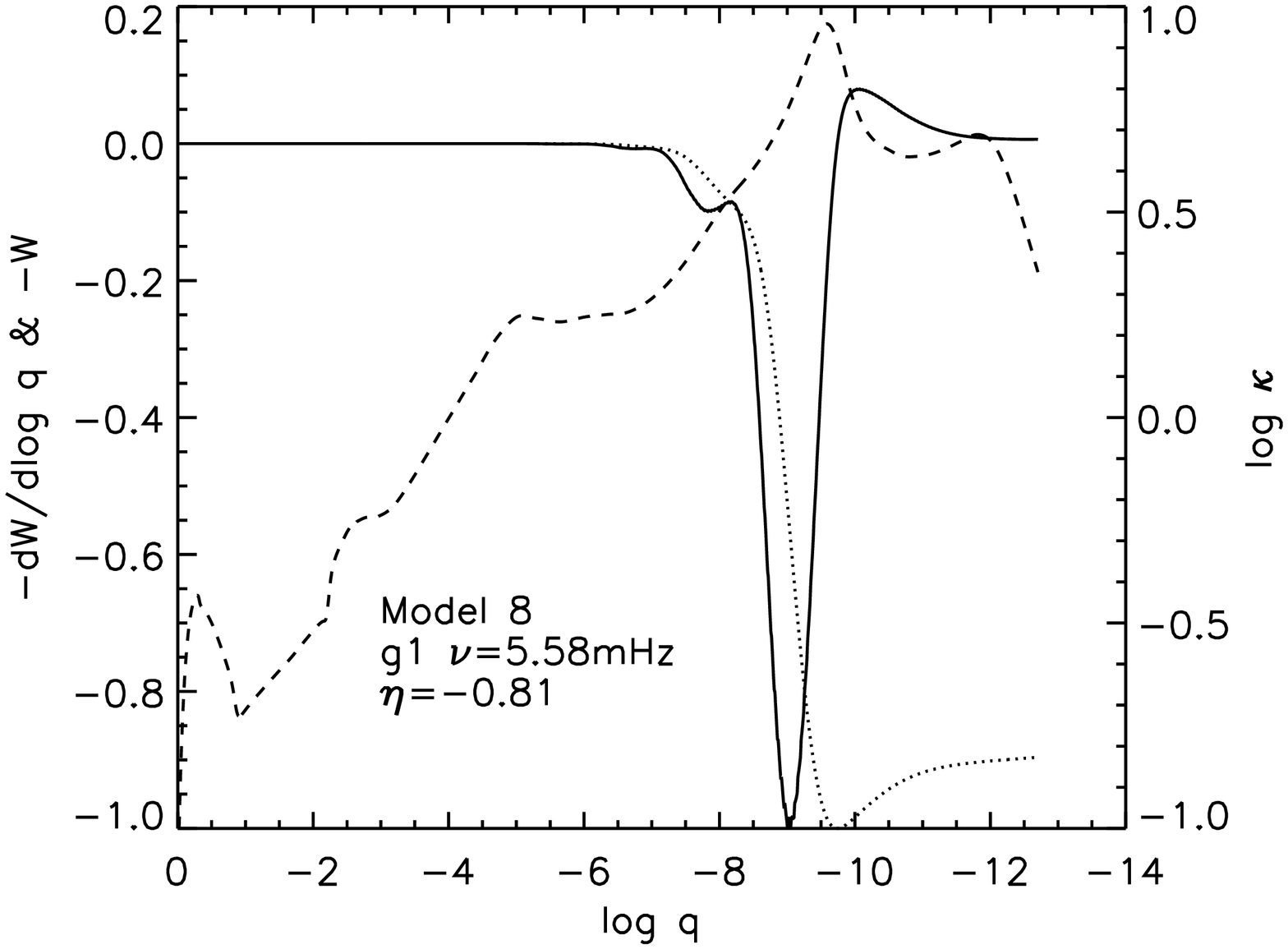}} \\
   \end{tabular}
   \caption{Left, Centre \& Right: Differential work (solid line), running work integral (dotted line) and opacity (dashed line) for the {\em g100}, {\em g15} and {\em g1} mode, respectively for a quadrupole mode of model 8. All plots have been scaled to arbitrary units.}
   \label{fig:modeseta650t45}
   \end{figure*}

   \begin{figure*}
   \begin{tabular}{cl}
   \resizebox{0.32\linewidth}{1.5in}{\includegraphics[angle=90]{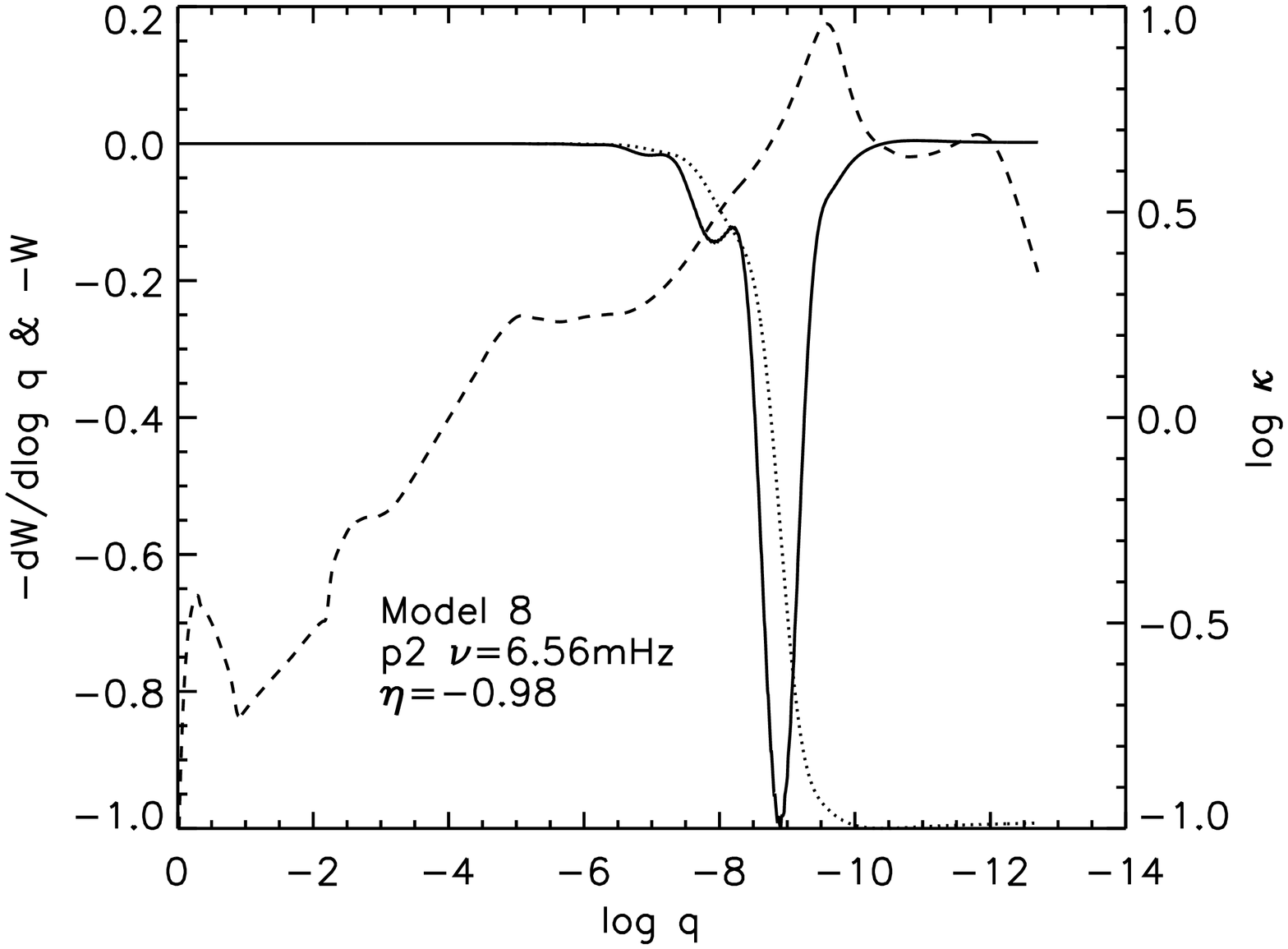}} 
   \resizebox{0.32\linewidth}{1.5in}{\includegraphics[angle=90]{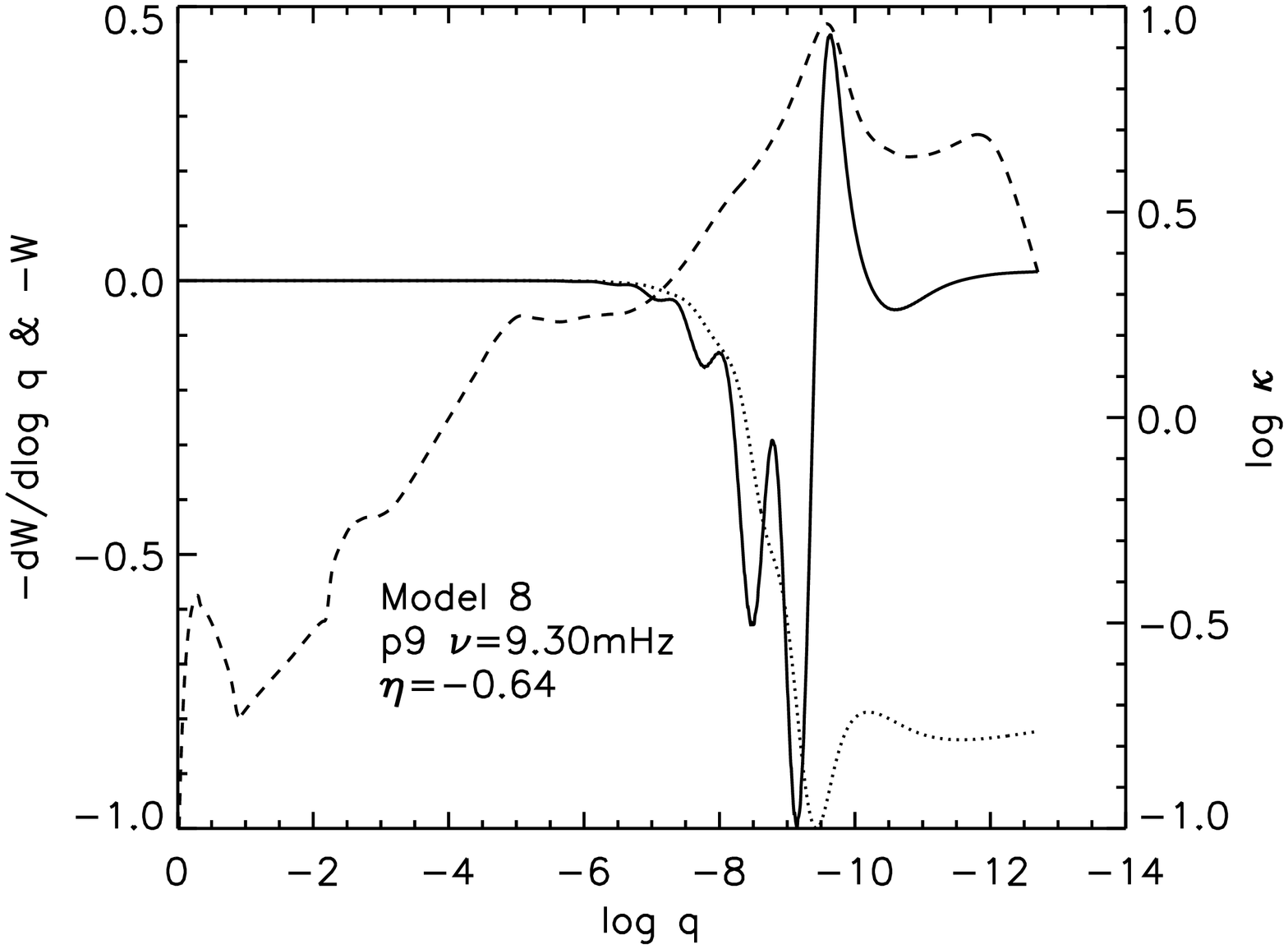}} &
   \resizebox{0.32\linewidth}{1.5in}{\includegraphics[angle=90]{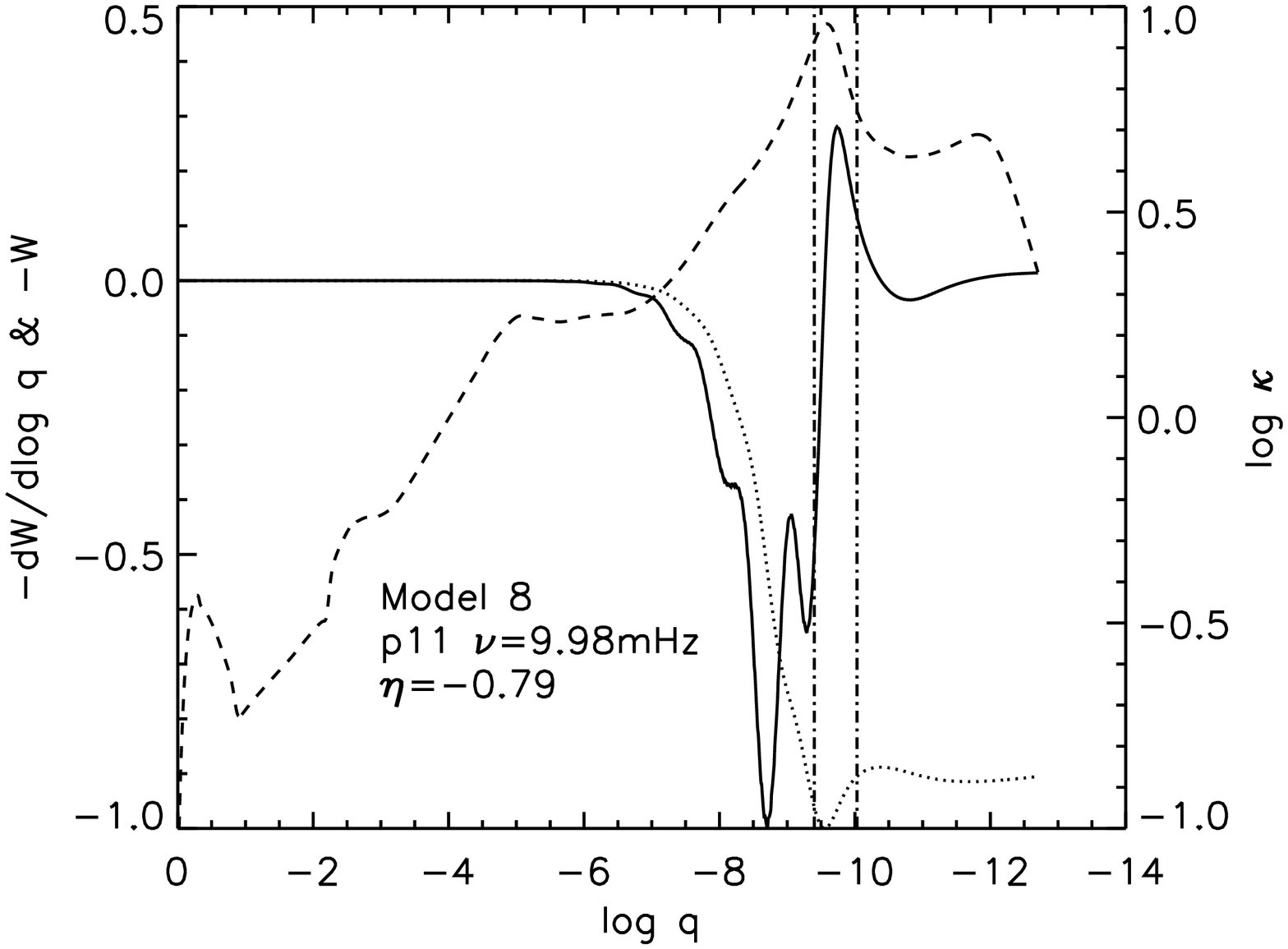}} \\
   \end{tabular}
   \caption{Left, Centre \& Right: Differential work (solid line), running work integral (dotted line) and opacity (dashed line) for the {\em p2}, {\em p9} and {\em p11} mode, respectively for a quadrupole mode of model 8. All plots have been scaled to arbitrary units. The vertical dashed-dotted lines in the last plot depict the convection zone.}
   \label{fig:modeseta650t45-2}
\end{figure*}

The model propagation diagram, representative of all the models used in the analyses, plus the opacity are shown in Fig.~\ref{fig:propaga}. There are three main bumps in the Rosseland opacity profile (scale given in the right axis): the shallowest at $\log q \simeq -12$ corresponds to the helium ionization zone, where He~II turns to He~III at $\sim 40\,000$~K; the maximum at $\log q \simeq -9.5$, which we will refer to as the Z-bump, is mainly due to iron-peak elements at $\sim 200\,000$~K; and the one at $\log q \simeq -5$ is the deep opacity bump at $\sim 2\,000\,000$~K, which results from a combination of K-shell photoionization of carbon and oxygen and spin–orbit effects in L-shell bound–free transitions of iron \citep{rogers92}. It is clear from the opacity profile that the He~II partial ionization zone is too high in the envelope to have any significant weight in the driving of the modes.

The Brunt-V\"ais\"al\"a ($\log N^2$) and the Lamb frequencies ($\log L_l^2$) are plotted for $l=2$. The Brunt-V\"ais\"al\"a frequency has maxima where gradients in chemical composition are steepest. Hence, the peak at $\log q \simeq -0.84$ accounts for the transition between the C/O core and the He burning shell, while the secondary peak at $\log q \simeq -2.23$ is the transition between the radiative He and H burning shell. All the energy transport in the model is radiative except at $-9.4 \leq \log q \leq -10.0$ where $N^2 < 0$, indicating that this is a convection zone associated with ionization of heavy elements producing an increase in the opacity rendering convective energy transport more efficient than radiation. We note here that models of sdB stars usually have a convective core and a radiative envelope with a thin convection zone produced by the partial ionization of HeII/HeIII {\citep{charpinet00}}.

   \begin{figure}
   \includegraphics[width=8.7cm]{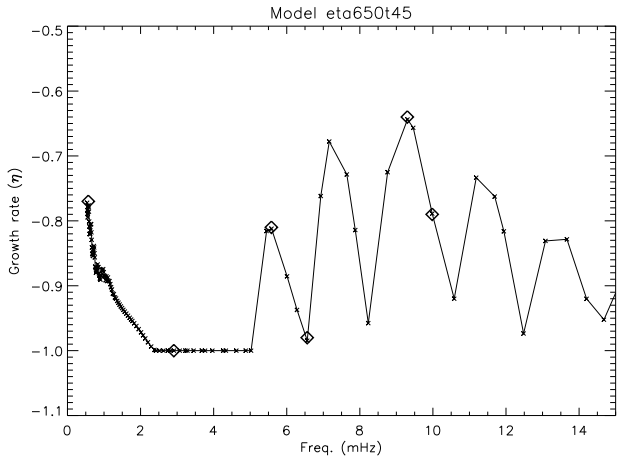} 
   \caption{Growth rate vs. frequency for $l=2$ modes of model~8. Diamonds indicate the modes for which the differential work and running work integral were plotted.}
   \label{fig:groueta650t45}
   \end{figure}

The {\em g}- and  {\em p}-mode propagation zones are determined respectively by $\sigma^2 < N^2, L_l^2$ and $\sigma^2 > N^2, L_l^2$, with $\sigma$ being the real part of the eigenfrequency. Fig.~\ref{fig:propaga} shows that {\em g}-modes cannot propagate in the outer envelope, and that {\em p}-modes cannot propagate in the innermost part of the star, except for modes with very high-radial order\footnote{The radial order, represented by $k$, or sometimes $n$, gives the number of nodes of each mode, in the radial direction.}. The diagram also gives broad constraints on the range of frequencies in which modes can propagate. It gives an upper limit for the angular frequencies of {\em g}-modes, $\log \sigma^2$ $\lesssim -1$ -- depending somewhat on the degree $l$ -- corresponding to cyclic frequencies $\nu \lesssim 50$~mHz or periods $P \gtrsim 20$~s. In the case of {\em p}-modes it gives a lower limit for eigenfrequencies, $\log \sigma^2 \gtrsim -4.3$, which translates to cyclic frequencies $\nu \gtrsim 1.1$~mHz or periods $P \lesssim 890$~s. Hence modes with oscillation eigenfrequencies in the range $-4.3 \lesssim \log \sigma^2 \lesssim -1$ are mixed modes, i.e. they behave as {\em p}-modes in the outer part of the star and as {\em g}-modes in the inner zones. These limits correspond to a range in the frequency propagation of $1.1 \lesssim \nu \lesssim 50$~mHz or a range in periods of $20 \lesssim P \lesssim -890$~s. These mixed modes are typical of evolved stars, like sdOs, and hamper the identification of the modes: P and G modes' propagation zones are not separated by an evanescent zone, but their frequency ranges overlap significantly, as derived above. We should then be cautious about the nomenclature and emphasise that whenever we refer to a {\em p}- ({\em g}-) mode, we refer in fact to mixed modes with nodes mainly in the P (G) mode propagation zone (see position of nodes for modes {\em p1}, {\em p30} in Fig.~\ref{fig:propaga}); except for those with high-radial orders (see nodes for mode {\em g88} in Fig.~\ref{fig:propaga}) i.e. {\em g}-modes with $\nu \lesssim 1.1$~mHz, and {\em p}-modes with $\nu \gtrsim 50$~mHz, far out of the frequency range considered here. In practice, for this model, modes with $\nu \gtrsim 22$~mHz (or k$>$33) do not have modes in the G mode region.

We plot the growth rate $\eta$ vs. frequency for $l=2$ modes (the behaviour being analogous for any degree $l$) in Fig.~\ref{fig:groueta650t45}, up to 15~mHz, as higher frequencies were found highly stable. The model is stable as the growth rate is negative in the whole frequency range. However there are 3 regions of interest in regard to stability:

\begin{itemize}
\item between about 0.5 to 1~mHz, where modes are not completely stable
\item from about 2.5 to 5~mHz, a highly-stable region
\item from $\sim5$~mHz to 15~mHz, where the growth rate shows an oscillatory behaviour. 
\end{itemize}

The differential work ($-dW/d \log q$), running work integral ($-W$), and opacity, for six modes (marked with diamonds in Fig.~\ref{fig:groueta650t45}) chosen to be representative of the growth rate profile, give some insight into the local contributions and their relative importance to the work integral.

Fig.~\ref{fig:modeseta650t45} shows the {\em g100} mode (on the left) with $\nu = 0.57$~mHz and $\eta = -0.78$, which is representative of the low frequency region where the growth rate is not so negative. We identify a driving region at the location of the deep opacity bump as responsible for the not complete stabilization of the modes. However, due to the large number of nodes in the deepest region, radiative damping exceeds driving and the modes are stable.

The {\em g15} mode (centre) with $\nu = 2.91$~mHz and $\eta = -1$ represents the high-stability region found at frequencies between 2.5--5~mHz. The region where maximum energy interchange for the model occurs where the ratio of the dynamical to thermal timescales is $\sim$ 1 (corresponding to $\log q \simeq -10$). For low-radial-order {\em g}-modes, maximum energy interchange takes place at $\log q \simeq -9$, where the derivative of the opacity does not favour driving, yielding a complete stabilization of the modes.

The onset of the oscillatory behaviour is represented by the {\em g1} mode (right) with $\nu = 5.58$~mHz and $\eta = -0.81$. The mode gains a bit in instability due to the development of a weak driving region at $-9.5 \lesssim \log q \lesssim -10$, i.e. the Z-bump location. Its oscillating nature is due to the rise and fall of a driving zone at this Z-bump location, as we show next.

We plot 3 modes belonging to the third region of stability in Fig.~\ref{fig:modeseta650t45-2}. On the left panel, we plot {\em p2} mode with $\nu = 6.56$~mHz and $\eta = -0.98$. We see that the reason for the low value of the growth rate is the vanishing driving region, whereas all the remaining energy contributes to damping.

The mode with maximum growth rate is {\em p9} (centre) with $\nu = 9.30$~mHz and $\eta_\mathrm{max} = -0.64$. The reason is the maximum developing of the driving region at the location of the Z-bump.

The decrease of the growth rate at high frequencies is due to the energy damping region becoming wider and wider, as is the case for the {\em p11} mode (right) with $\nu = 9.98$~mHz and $\eta = -0.79$.

Thus, the interplay between the development and extinction of a driving region at the Z-bump location is responsible for the oscillatory behaviour of the growth rate.

In an attempt to thoroughly investigate the properties of this model (and in particular the peculiar behaviour of the growth rate), we explored the behaviour of the kinetic energy of the modes. This lead to the discovery of a mode trapping structure in the {\em g}- and {\em p}-mode spectra, the details of which will be presented elsewhere \citep{crl09b}.

\section{Model 10: eta\,675t45mixi}

This is the first of seven models -- models 10 to 16, see Table~\ref{tab:models}-- created by introducing an artificial extra-mixing in the post-ZAHB evolution with the purpose of raising the metallicity in the envelope in an attempt to drive pulsations, as explained in Section~2.1.

Model 10 was our first model discovered to be unstable. We found that high-radial order {\em g}-modes are driven through the action of a $\kappa$-mechanism due to the Z-bump at $\sim 200,000$~K, the same mechanism that drives oscillations in sdB stars. Although the extra-mixing has significantly increased the abundances of C and O in the driving region, this has very little effect on the opacity there. Hence the presence of instability in this model shows that radiative levitation of iron is not a necessary condition for pulsations of sdO stars, as long as the metallicity is sufficiently increased. More evolved models from this evolutionary sequence (models 10.1 - 10.6 in Table~\ref{tab:models}) were also tested for pulsational instability. We found pulsations were driven in models with $T_\mathrm{eff} \leq 54\,000$~K and $\log g \leq 4.5$. So, the instability strip for this evolutionary sequence was found to be $45\,000 \leq T_\mathrm{eff}\leq 54\,000$~K and $4.2 \leq \log g \leq 4.5$, and may extend to lower temperature and surface gravity.

\begin{figure}
   \includegraphics[width=8.5cm]{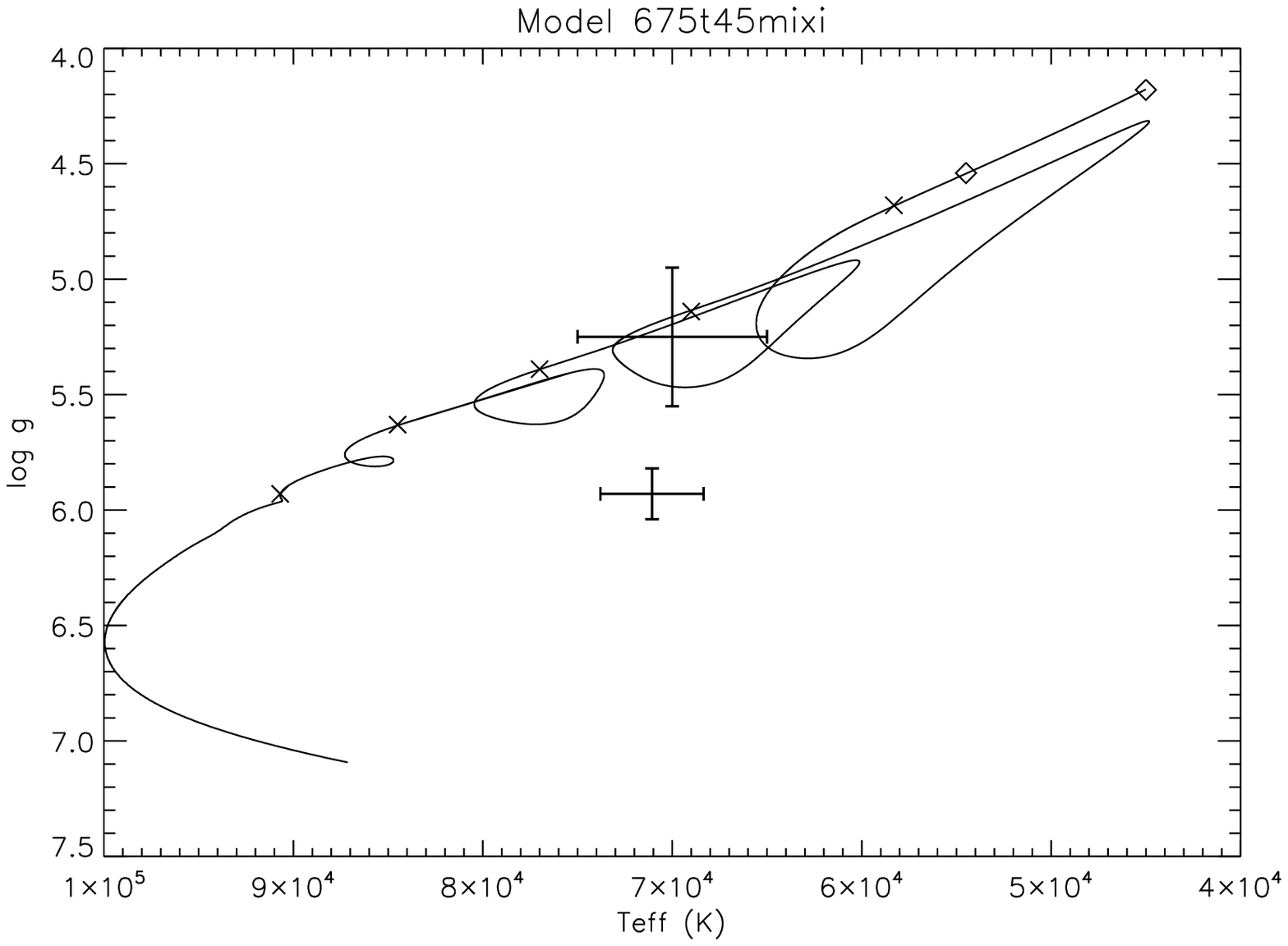} 
   \caption{Evolutionary track described by model 10 (initial diamond). Diamonds mark models found unstable and crosses those found stable. Error bars give the $T_\mathrm{eff}$ and $\log g$ determination for J\,1600+0748 by \citet{fontaine08} (bottom) and \citet{crl09a} (top). Extra mixing in the model carries helium to hotter regions were it burns giving rise to thermal pulses. The energy produced causes the star to expand and cool evolving to lower temperatures and when the burning finishes the star returns to higher temperatures.}
   \label{fig:dhrmodel10}
   \end{figure}

The evolutionary track on the $\log g$--$T_\mathrm{eff}$ diagram described by the time evolution of the model is shown in Fig.~\ref{fig:dhrmodel10}. The positions of the initial model 10 and the subsequent model found unstable are marked with diamonds. Crosses indicate intermediate structural models found stable in the stability analysis. The whole evolutionary time for this model is about 300\,000~yr, while e-folding times of the excited modes are of the order of $10^{-1}$ to 10~yr, so modes have enough time to grow in amplitude.

\begin{figure}
   \includegraphics[width=8.5cm]{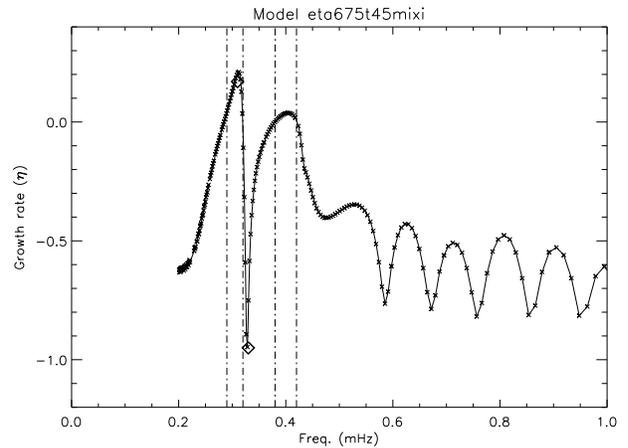} 
   \caption{Growth rate vs. frequency for model 10. Diamonds indicate the modes for which the differential work and running work integral were plotted. The dashed-dotted lines delimit the two region of unstable modes. Higher frequency modes where found highly stable and were not plotted for clarity reasons.}
   \label{fig:grou675t45mixi}
   \end{figure}

The growth rate profile for $l=2$ modes is shown in Fig.~\ref{fig:grou675t45mixi}. We found two regions with excited modes (similar growth rate profiles and instability regions where found for the other degrees): 

\begin{itemize}
\item within the frequency range $0.29 \leq \nu \leq 0.32$~mHz (periods 52--58~min), which correspond to modes from {\em g203} to {\em g182} and
\item within the frequency range $0.38 \leq \nu \leq 0.42$~mHz (periods 40--44~min), which correspond to modes from {\em g152} to {\em g138}
\end{itemize}


   \begin{figure*}
   \begin{tabular}{cl}
      \resizebox{0.32\linewidth}{1.5in}{\includegraphics[angle=90]{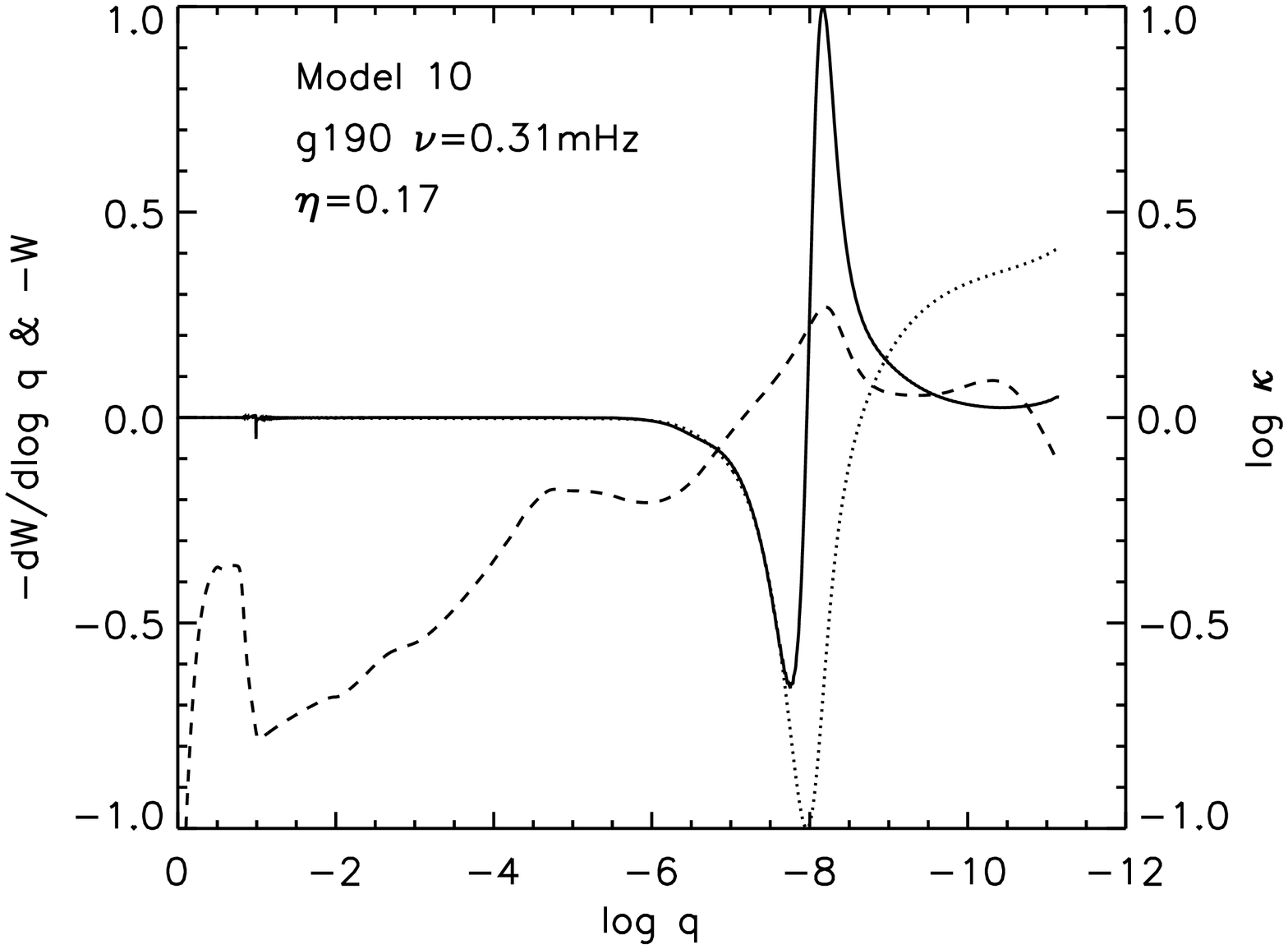}} 
      \resizebox{0.32\linewidth}{1.5in}{\includegraphics[angle=90]{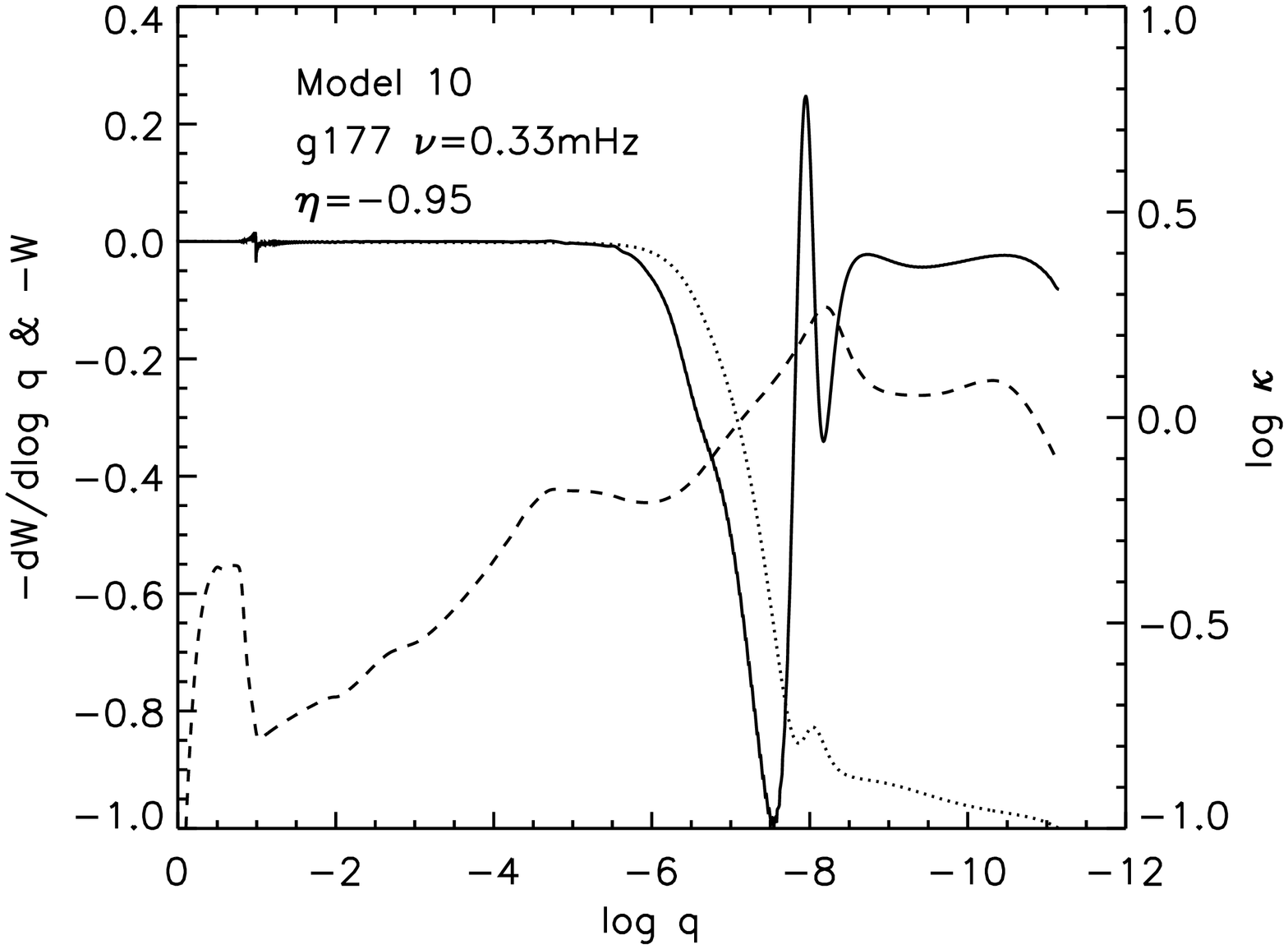}} &
      \resizebox{0.32\linewidth}{1.5in}{\includegraphics[angle=90]{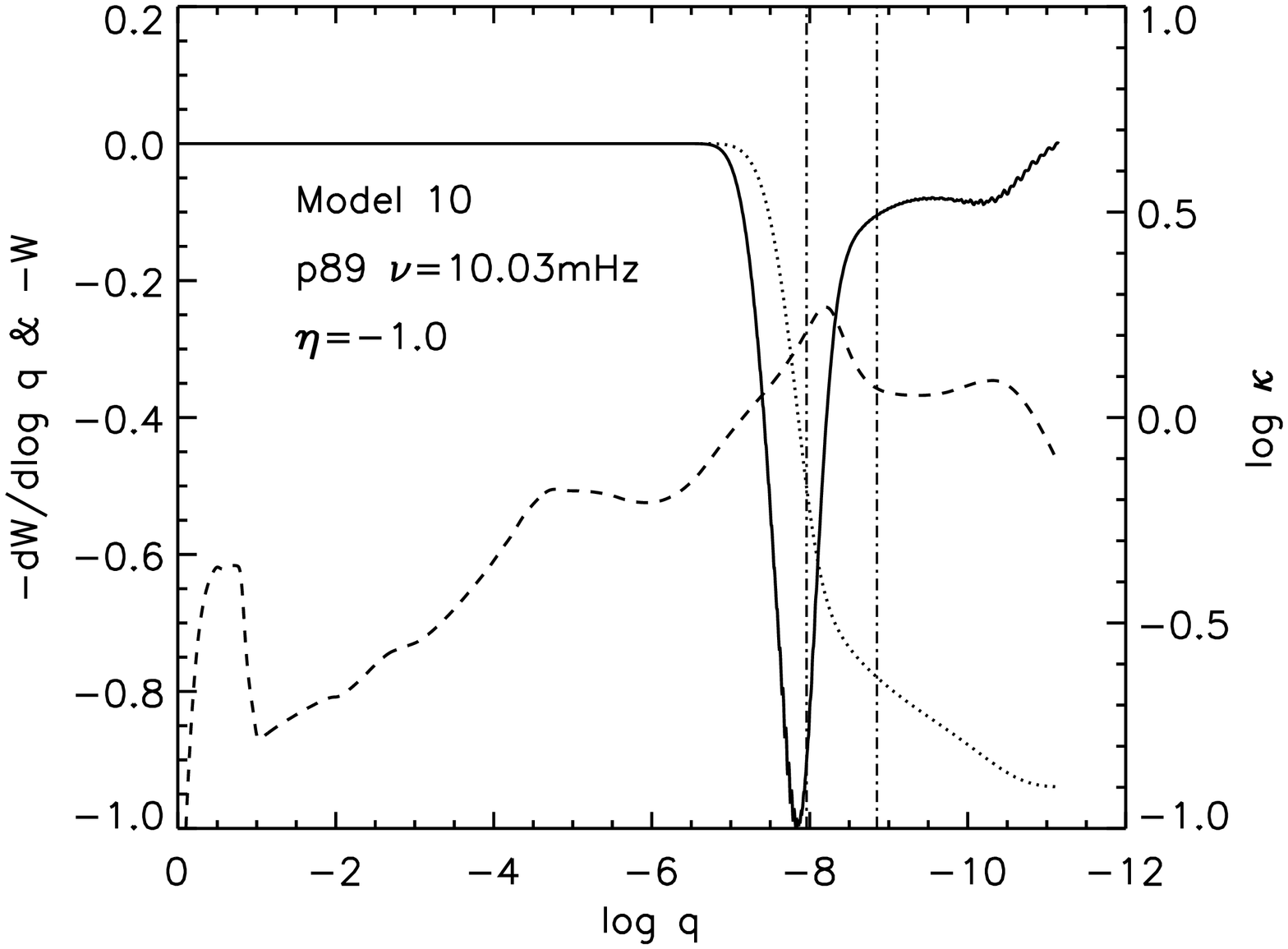}} \\
   \end{tabular}
   \caption{Left, Centre \& Right: Differential work (solid line), running work integral (dotted line) and opacity (dashed line) for the {\em g190}, {\em g177} and {\em p89} mode, respectively for a quadrupole mode of model 10. All plots have been scaled to arbitrary units. The vertical dashed-dotted lines in the right plot depict the convection zone.}
   \label{fig:modes675t45mixi}
   \end{figure*}

These modes fall within a wider range favoured for instability between $0.2 \lesssim \nu \lesssim 1$~mHz. Meanwhile, modes with frequencies $\nu \gtrsim 2$~mHz were found highly stable with $\eta = -1.0$. We have selected a few representative modes, marked with diamonds in Fig.~\ref{fig:grou675t45mixi}, for further discussion. The profiles of the differential work, running work integral, and opacity for these representative modes are shown in Fig.~\ref{fig:modes675t45mixi}.

We chose the {\em g190} mode with $\nu = 0.31$~mHz and $\eta= 0.17$ (Fig.~\ref{fig:modes675t45mixi}, left) as representative of the two `islands' of instability. There is a wide driving zone centered at $\log q \simeq -8$, the location of the Z-bump responsible for the instability.

To represent the stability zone between the two `islands' of instability and, moreover, the stability zone up to 2~mHz, we chose the {\em g177} mode (Fig.~\ref{fig:modes675t45mixi}, centre) with $\nu = 0.33$~mHz and $\eta= -0.95$. The stability is due to the near extinction of the previous driving region and most of the significant energy contributing to damp the modes. The development and extinction of the region contributing to driving (with its maximum taking place at the location of the Z-bump) at the expense of a damping region is produced in the frequency range between $\sim$ 0.2 to 1.5~mHz and is responsible for the oscillatory profile of the growth rate.

Finally, mode {\em p89} (Fig.~\ref{fig:modes675t45mixi}, right) with $\nu = 10.03$~mHz and $\eta= -1.0$ is representative of the high stability of low and high-radial order {\em p}-modes. The stability is due to a wide damping region at the location of the Z-bump.

Model 10.1 of this same evolutionary sequence was found unstable in the frequency range (for $l=2$, and similarly for other degrees):
\begin{itemize}
\item 0.71 to 0.75~mHz (periods 22-23~min) which corresponds to modes from {\em g}81 to {\em g}76.
\end{itemize}

\section{Model 15: eta\,675t90mixi}

\begin{figure}
   \includegraphics[width=8.5cm,angle=-90]{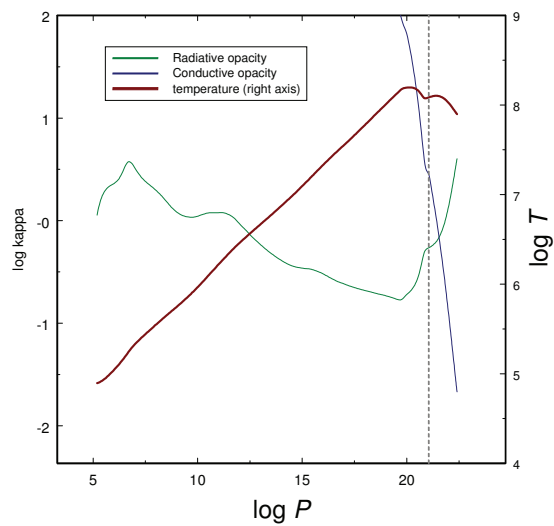} 
   \caption{Radiative (green) and conductive (blue) opacity vs. pressure. Temperature (maroon) is given in the right axis. The vertical dashed line marks the opacity maximum. See text for details.}
   \label{fig:opacity}
   \end{figure}

This is the second model found unstable in our study. We found it drives two overstable modes. This driving is achieved through the action of three different driving regions: (i) the opacity bump at $\sim 2,000,000$~K which for this model is due to a combination of opacity from iron-peak elements and, because of their increased abundances due the extra-mixing, opacity from C and O, (ii) a driving region at $\sim 2 ~10^7$~K, located immediately above the hydrogen burning shell, and most likely due to the K-edge of highly ionised argon \citep{rogers92}, and (iii) an opacity bump that occurs at the transition from radiative to conductive energy transport. At the high temperatures and densities in the core the radiative opacity is dominated by free - free absorption. Because the core temperature gradient is relatively small, the radiative opacity increases with the density. On the other hand the electrons become more degenerate with increasing density and the conductive opacity decreases. The combination of these two opacity variations gives a maximum in the opacity, corresponding to the opacity bump in driving region (iii) (see Fig.~\ref{fig:opacity}). Tests cancelling out the contribution of the nuclear energy generation show that the $\epsilon$ mechanism is not responsible for this deepest driving region, nor for region (ii). The instability was obtained for $l=2$ modes (similar for other modes with other degrees): 

\begin{itemize}
\item {\em g27} with $\nu=2.08$~mHz and $\eta=0.07$ and 
\item {\em g24} with $\nu=2.32$~mHz and $\eta=0.05$.
\end{itemize}

This corresponds to periods in the range 7--8 min. Model~15 has one of the lowest H mass fraction, and one of the largest He, C and O mass fractions (see Table~\ref{tab:models}), which extend with constant values from the surface of the star down in the envelope until about $\log q \simeq -4$. It has also the largest metallicity, Z=0.19, of the whole series of models.

\begin{figure}
\centering 
\includegraphics[width=8.5cm]{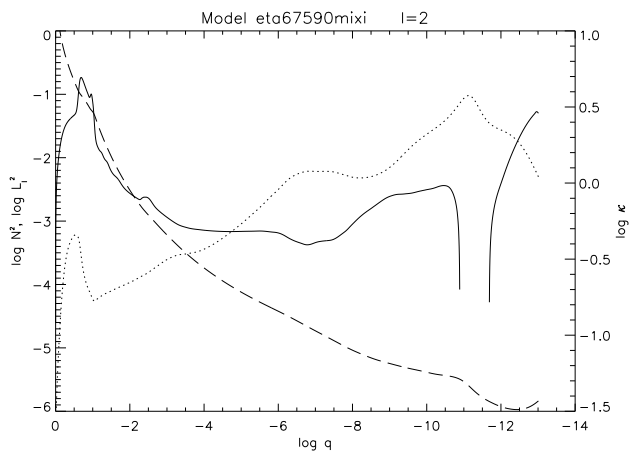} 
\protect 
\caption{Brunt-V\"ais\"al\"a (solid line), Lamb frequency for $l=2$ and opacity (dashed line) for model 15. Both in units of $\sigma^2$.}
\label{fig:phys3model15}
\end{figure}

The propagation diagram and opacity of model 15 (Fig.~\ref{fig:phys3model15}) shows that the region of mixed modes is reduced compared to that of model 8, which is representative of most of the models used in this study. We also note that the Z-bump has shifted to shallower regions, with its maximum at $\log q \simeq -11$ compared to $\log q \simeq -10$ for the models with $T_\mathrm{eff}=70\,000$~K or $\log q \simeq -8$ for models with $T_\mathrm{eff}=45\,000$~K.

   \begin{figure}
   \includegraphics[width=8.5cm]{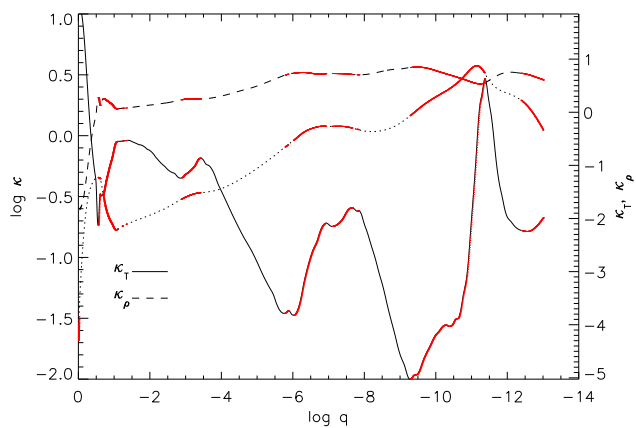} 
   \caption{Rosseland mean opacity $\kappa$ (dotted line, left axis), and its opacity derivatives (right axis): $\kappa_T$ (solid line) and $\kappa_{\rho}$ (dashed line). Red regions fulfil the criterion for instability. See text for details.}
   \label{fig:kappas}
   \end{figure}

\begin{figure*}
\centering
\begin{tabular}{cl}
\resizebox{0.32\linewidth}{1.5in}{\includegraphics[angle=90]{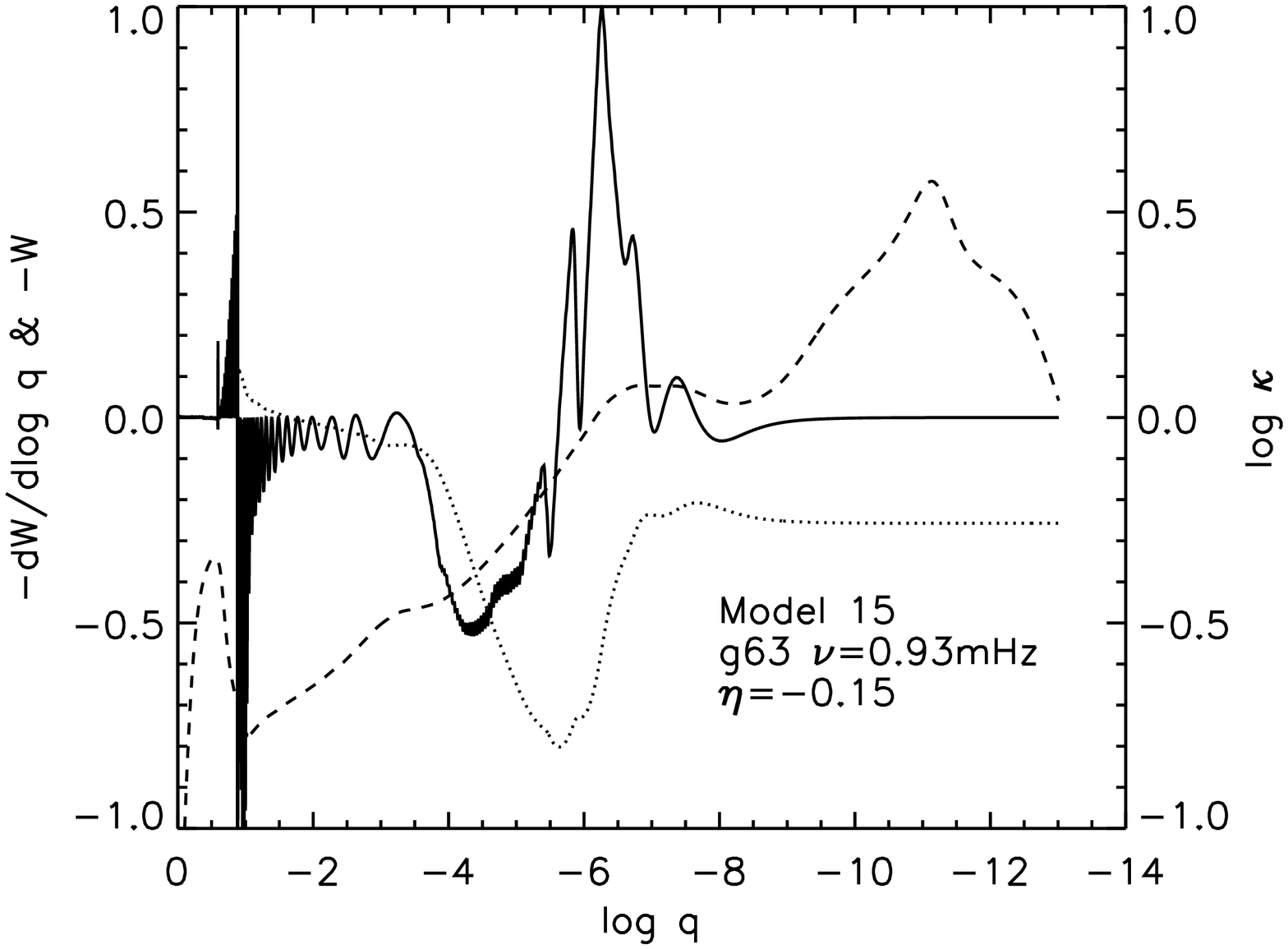}} 
\resizebox{0.32\linewidth}{1.5in}{\includegraphics[angle=90]{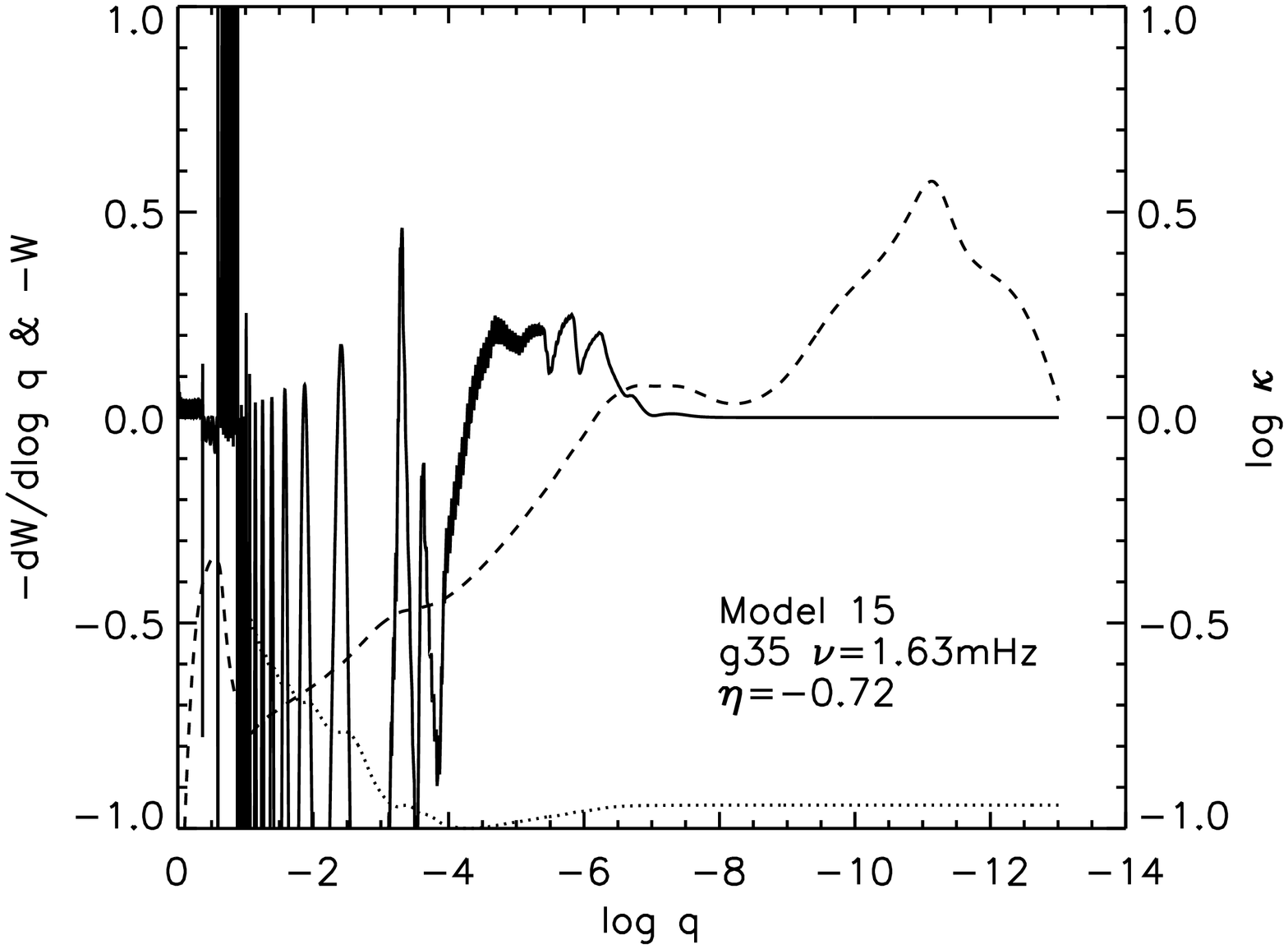}} &
\resizebox{0.32\linewidth}{1.5in}{\includegraphics[angle=90]{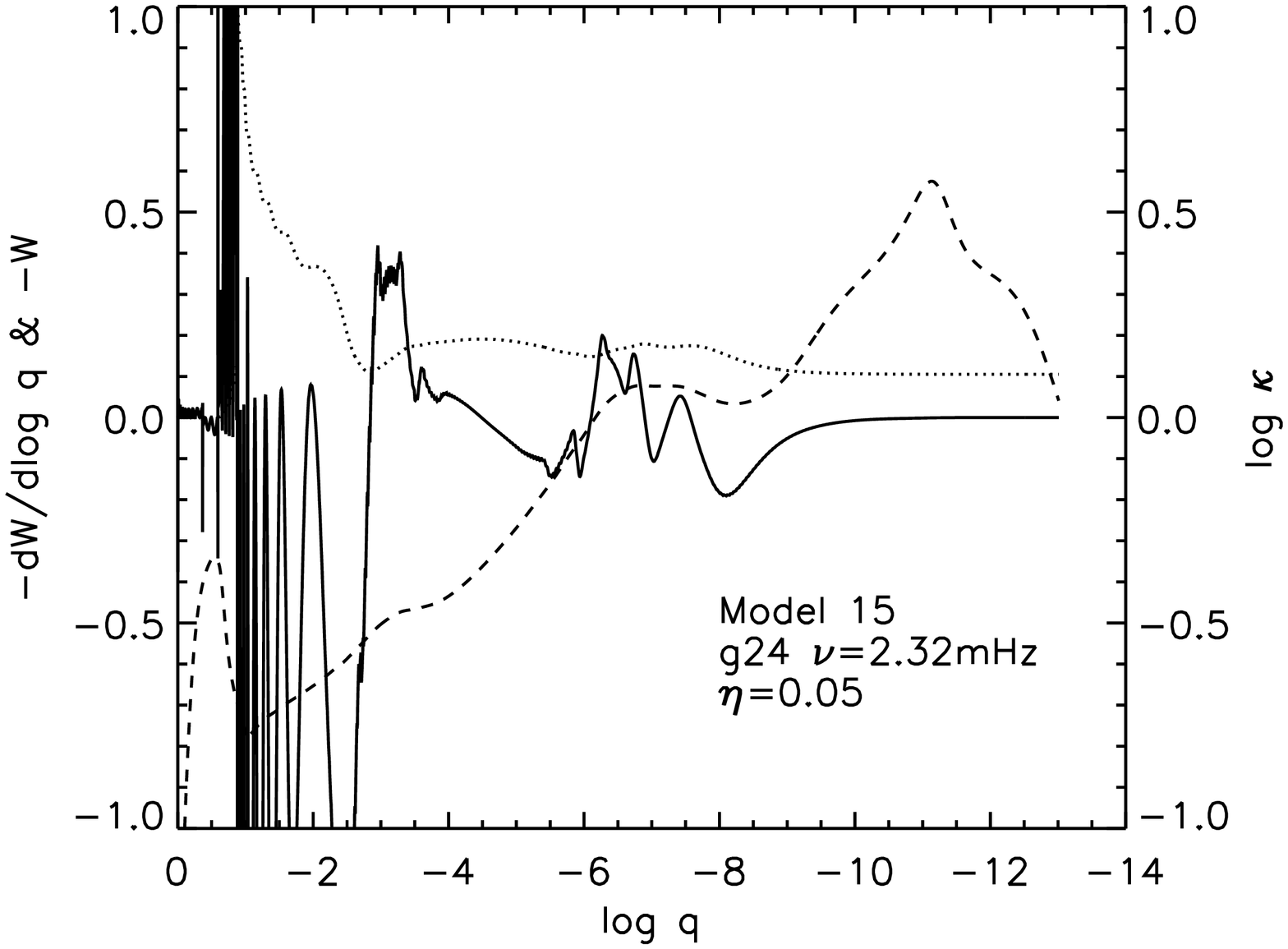}} \\
\protect
\end{tabular}
\caption{Left, Centre \& Right: Differential work (solid line), running work integral (dotted line) and opacity (dashed line) for the {\em g63}, {\em g35} and {\em g24} mode respectively for a quadrupole mode of model 15. All plots have been scaled to arbitrary units.}
\label{fig:modes675t90mixi-1}
\end{figure*}

\begin{figure*}
\centering
\begin{tabular}{cl}
\resizebox{0.32\linewidth}{1.5in}{\includegraphics[angle=90]{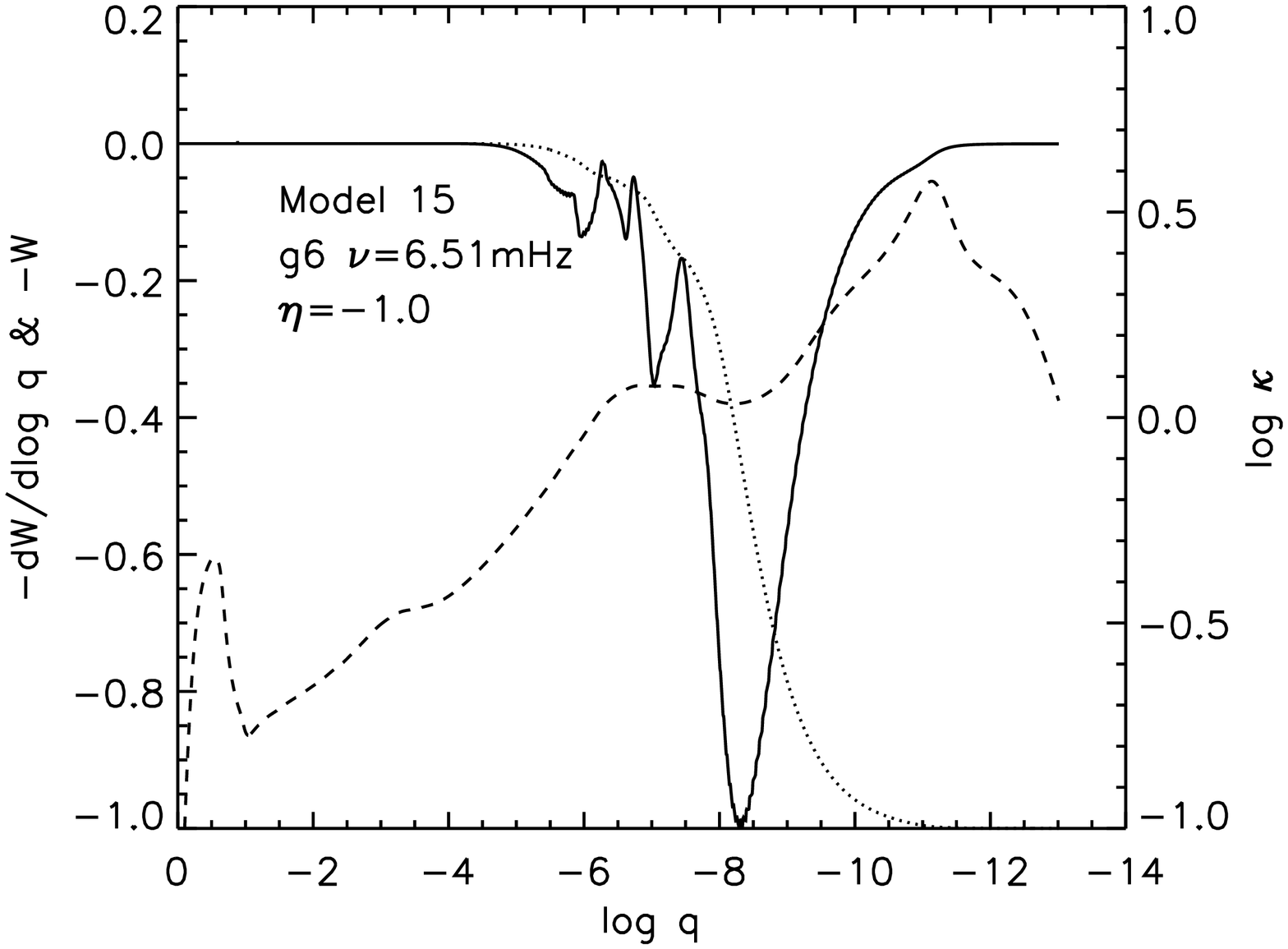}} &
\resizebox{0.32\linewidth}{1.5in}{\includegraphics[angle=90]{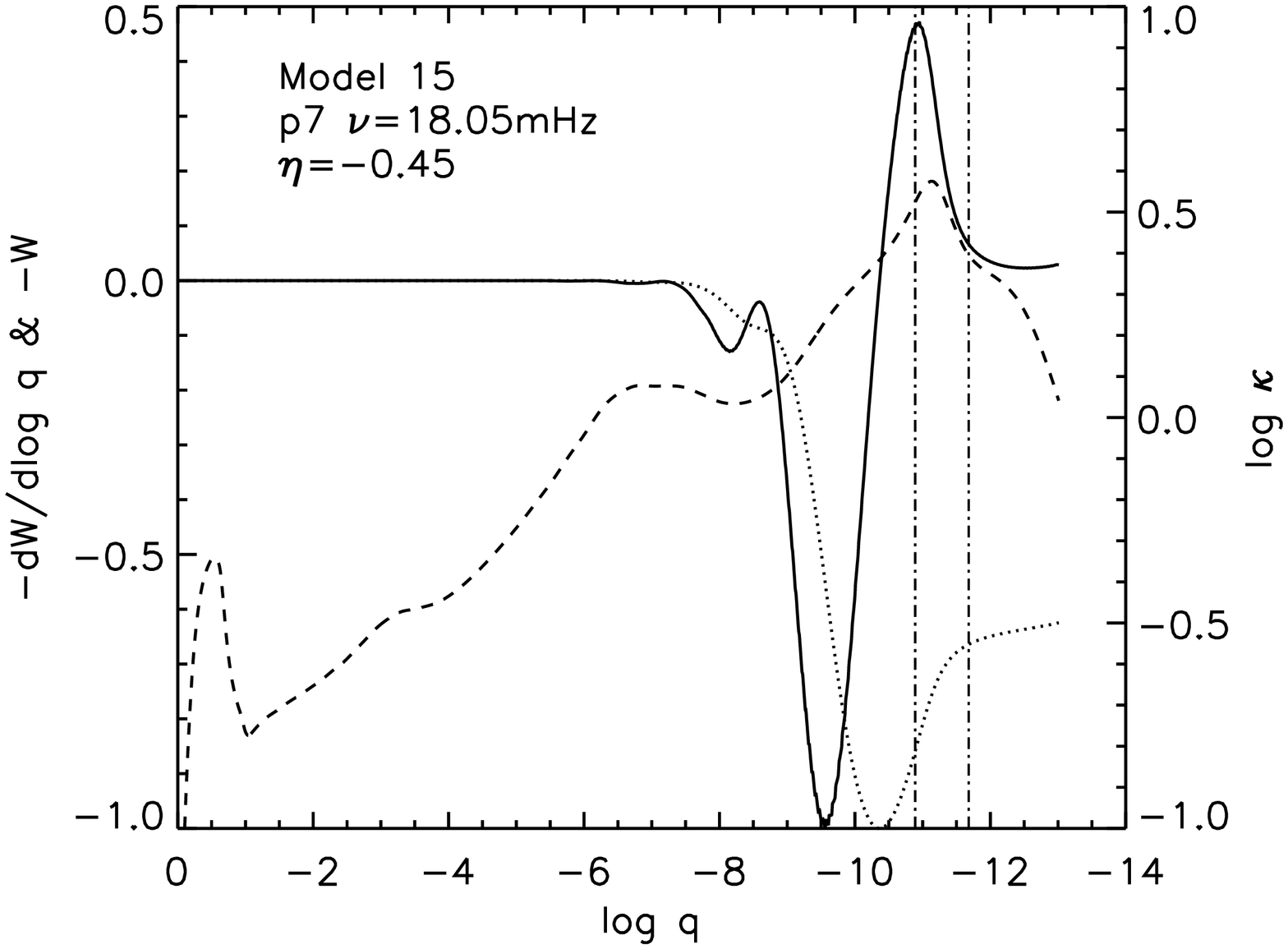}} \\
\protect
\end{tabular}
\caption{Left, \& Right: Differential work integral (solid line), running work integral (dotted line) and opacity (dashed line) for the {\em g6} and {\em p7} mode respectively for a quadrupole mode of model 15. The vertical dashed-dotted lines in the right plot depict the convection zone. All plots have been scaled to arbitrary units.}
\label{fig:modes675t90mixi-2}
\end{figure*}

A plot of the Rosseland mean opacity and its opacity derivatives, $\kappa_T = \partial \ln \kappa / \partial \ln T|_{\rho}$ and $\kappa_{\rho}= \partial \ln \kappa / \partial \ln \rho |_T$ is shown in Fig.~\ref{fig:kappas}. Opacity derivatives play an important role in the $\kappa$ mechanism as the necessary condition for driving is given by the expression \citep{unno89}:

\begin{equation}
 \frac{d}{dr} \left( \kappa_T + \frac{\kappa_{\rho}}{\Gamma_3 -1} \right) > 0
\end{equation}

\noindent Regions in the star fulfilling this criterion are shown in red in Fig.~\ref{fig:kappas}. All the regions involved in the effective driving of the two excited modes are potential driving regions, together with the region of the Z-bump, also involved in the {\em p}-mode excitation (see below) and the very surface of the star, which, on the contrary, has no significant weight in the driving.

The age of this model from when it first became an sdO is about 10\,000 yr, which is less than the linear growth rate e-folding times found: $7\,10^4$ and $10^5$~yr for {\em g}24 and {\em g}27, respectively. However, the effective temperature remains near 90\,000~K for more than $10^5$~yr after this particular model. Therefore, from the rate of change of the effective temperature, the evolutionary time scale for this model is about $10^5$~yrs, and modes may be able to develope observable amplitudes.

\begin{figure}
\centering 
\includegraphics[width=8.5cm]{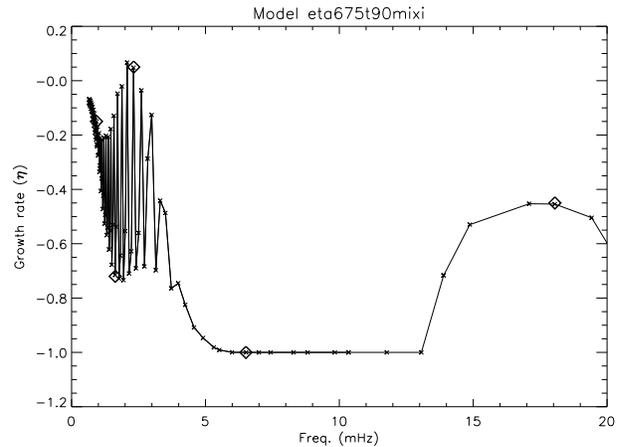} 
\protect 
\caption{Growth rate vs. frequency for model 15. Diamonds indicate the modes for which the differential work and running work integral were plotted.}
\label{fig:grou675t90mixi}
\end{figure}

The normalised growth rate of the model is given in Fig.~\ref{fig:grou675t90mixi}. It has a low frequency region with $0.5 \lesssim \nu \lesssim 3.0$~mHz where the two overstable modes are driven, and in which high and low values of the growth rate alternate for consecutive modes. There is a highly stable region at intermediate frequencies $5.0 \lesssim \nu \lesssim 13.0$~mHz and finally, a tendency to driving is found in the high frequency range with $14.0 \lesssim \nu \lesssim 20$~mHz where low-radial order {\em p}-modes are found.

An example of a quasi-unstable mode in the very low frequency region, below 1~mHz, is given in Fig.~\ref{fig:modes675t90mixi-1} (left) by the {\em g63} mode with $\nu=0.93$~mHz and $\eta=-0.15$. The high values of the growth rate are due to driving from region (i). At the high effective temperatures of this model this opacity bump occurs high in the envelope, at $\log q \simeq -6$, which favours pulsations, as heat interchange is more effective.

An example of a mode in the low frequency region which has lost its tendency to driving is given in Fig.~\ref{fig:modes675t90mixi-1} (centre) by the {\em g35} mode with $\nu=1.63$~mHz and $\eta=-0.72$. In addition to driving region (i), a driving region at $-1 \lesssim \log q \lesssim -0.5$ (corresponding to $\sim 8~10^7$~K in temperature) has developed near the location of the deep opacity bump due to the transition from radiative to conductive energy transport in the semi-degenerate material. However, radiative damping right above this region exceeds the driving effect of both regions, damping the modes.

One of the two unstable modes found, {\em g24} with $\nu=2.32$~mHz and $\eta=0.05$ is shown in Fig.~\ref{fig:modes675t90mixi-1} (right). A new driving region, which we attribute to Ar ionization, has developed at $\log q \sim -3$, where $T \sim 210^7$~K. The model is able to achieve instability due to the combined driving of two of the three regions: the first and second, or the first and third regions, but modes are stable if only the first, or the joint second and third regions contribute to the differential work. It is the first time, to our knowledge, that driving of modes due to such a combined action of driving regions has been reported. This may give us a useful tool to explore physical processes, in particular energy transport, in dense hot regions.

A mode representative of the intermediate frequency region with the lowest values of the growth rate is the {\em g6} mode with $\nu=6.51$~mHz and $\eta=-1.0$ (Fig.~\ref{fig:modes675t90mixi-2}, left). Modes at intermediate frequencies have maximum energy interchange at $\log q \simeq -8.5$, where the derivative of the opacity does not favour driving, which results in maximum stabilization of the modes.

Fig.~\ref{fig:modes675t90mixi-2} (right) shows mode {\em p7} with $\nu=18.05$~mHz and $\eta=-0.45$, as an example of a quasi-unstable mode in the high frequency region where the low-radial order {\em p}-modes are found. This mode falls within the region of observed modes for J1600+0748. Surprisingly, due to the high $T_\mathrm{eff}$ of the model, the high values of the growth rate are due to a wide driving region caused by Z-bump. However, driving is not achieved as it is counterbalanced by a damping region located at the base of the Z-bump.

Further evolved models in the evolutionary sequence were analysed and found stable.

\section{Summary of the results}

We present a summary of the stability analysis carried out for all the analysed models and submodels in Table~\ref{tab:piz}. We give $T_\mathrm{eff}$, $\log g$, metallicity and if a tendency to instability -- or actual instability -- was found in the {\em g}- and/or {\em p}-mode spectrum and which driving region was responsible for it. In addition, we give the frequency and period range most favoured for instability.

\begin{table*}
\centering
\caption{Effective temperature, logarithm of surface gravity, current metallicity and region responsible for the tendency to driving ---or actual driving--- of the g- and/or p-modes spectrum. The last two columns give the frequency or period range of most favoured for driving (taking into account modes from $l=1$ to $l=4$). The lower-case z for models 7 and 8 indicates the modes are under the limit of the tendency to driving. Models between horizontal lines have been calculated with small changes in the evolution code. Models in bold were found unstable.}

\label{tab:piz}
\setlength{\tabcolsep}{4pt}
\begin{tabular}{l|cccccccccccccc}  
Model number &  T$_\mathrm{eff}$ & log g  &   Z   & g-modes & p-modes & Freq. range & Period range\\ 
\& name      &  (K)       &        &       &         &         &  (mHz)      &      (s) \\
\hline
\hline
1  p675\_8057 & 79\,000   & 5.70   & 0.07  &  C/O    &  Z   & 0.5--1.5, 10--14  & 700--2000, 70--100 \\
1.1  p675\_9065 & 90\,000 & 6.50   & 0.07  &  C/O    &  --  & 0.5--5   & 200--2000 \\
2  p650\_7960 & 79\,000   & 5.95   & 0.02  &  C/O    &  --  & 0.5--1   & 1000--2000\\
\hline						        	  	 
3  etap\,685  & 55\,000   & 5.89   & 0.02  &  C/O    &  --  & 0.5--2.5 & 400--2000 \\
4  etap\,690  & 55\,000   & 5.95   & 0.02  &  C/O    &  --  & 0.5--2.5 & 400--2000 \\
5  etap\,695  & 55\,200   & 5.98   & 0.02  &  C/O    &  --  & 0.5--3   & 300--2000 \\
6  etap\,700  & 55\,000   & 6.02   & 0.04  &  C/O    &  --  & 0.5--2.5 &  \\
\hline							       
7  eta\,600t45 & 45\,000  & 4.95  & 0.05  &   z     &  --  & 2--5     & 200--500  \\
7.1  eta\,600t70 & 70\,000 & 5.95  & 0.05  &  C/O    &  --  & 0.5--1.5 & 700--2000 \\
8  eta\,650t45 & 45\,000  & 5.26  & 0.05  &   --    &  z   & 7--9.5   & 100--150  \\
8.1 eta\,650t70 & 70\,000 & 6.22   & 0.05  &  C/O    &  --  & 0.5--3.5 & 300--2000 \\
9  eta\,700t45 & 45\,000  & 6.13   & 0.07  &  C/O    &  --   & 1--7    & 150--1000 \\
9.1 eta\,700t70 & 70\,000 & 6.34   & 0.07  &  C/O    &  --  & 0.5--4   & 250--2000 \\
\hline							       
{\bf 10 eta\,675t45mixi} & 45\,000 & 4.18 & 0.14 & Z &  --  & 0.2--0.9 & 1100--5000 \\
{\bf 10.1 15773t54g45}   & 54\,500 & 4.54 & 0.15 & Z &  --  & 0.3--2   & 500--3300 \\ 
10.2 15873t58g47 & 58\,300 & 4.68  & 0.14  &  Z      &  --  & 0.3--2.5 & 400--3000 \\ 
10.3 16973t69g51 & 69\,000 & 5.14  & 0.16  & C/O+Z   &  --  & 0.2--3.5 & 300--5000 \\ 
10.4 17773t77g54 & 77\,000 & 5.39  & 0.17  & C/O+ Z  &  --  & 0.2--1.5, 4--6.5 & 700--5000, 150--250 \\ 
10.5 18273t84g56 & 84\,500 & 5.63  & 0.18  &  C/O    &   Z  & 0.2--1, 8--11 & 1000--5000, 90--125 \\ 
10.6 18573t91g59  & 90\,700 & 5.93  & 0.19  &  C/O    &   Z  & 0.2--3, 17--21& 300--5000, 50--60 \\ 
11 eta\,675t45mixnmi & 45\,000 & 4.36 & 0.07 &   Z   &  --  & 0.4--0.7 & 1400--2500 \\
12 eta\,675t70mix1i  & 70\,000 & 5.16 & 0.16 &C/O + Z&  --  & 0.2--1, 2.5--4 & 1000--5000, 250--400 \\
13 eta\,675t70mix2i  & 70\,000 & 5.20 & 0.18 &C/O + Z&  --  & 0.2--0.5, 3--4 & 2000--5000, 250--300 \\
14 eta\,675t70mixnmi & 70\,000 & 5.43 & 0.07 &C/O + Z&  --  & 0.4--0.8, 5--7 & 1250--2500 \\
{\bf 15 eta\,675t90mixi} & 90\,000 & 5.86 & 0.19 &C/O&   Z  & 0.3--3.5, 15--20 & 300--3300, 50--65 \\
15.1 18742t98g63     & 97\,600 & 6.28 & 0.20 &  C/O  &  --  & 1.5--2 &  500--700 \\ 
16 eta\,675t90mixnmi & 90\,000 & 6.58 & 0.07 &  C/O  &  --  & 2--4, 5.5--7 & 250--500, 150--180 \\
\hline
\end{tabular}							       
\end{table*}

If we draw our attention first to the first batch of models built for this study (models 1 to 9), all of them, except models 7 and 8, have a certain tendency to instability of the {\em g}-modes due to the gathering of driving energy at the location of the deep opacity bump at $T \sim 2 ~10^6$~K. This opacity bump is mainly due to iron-peak elements in models which have not experienced mixing of the products of core helium burning into the envelope. Opacity from C and O plays a role for models in which mixing has occurred. Whether or not opacity from C and O is important for driving, we label this driving region C/O in Table~\ref{tab:piz}.  Modes more favoured for driving are high-radial order {\em g}-modes, {\em i.e.} long period modes, with typical frequencies of oscillation between $\sim$0.5 and 2~mHz, i.e. periods $\sim$500 to 2000~s. We emphasis the unique behaviour of model 7 in which the tendency to instability of the {\em g}-modes is caused by the Z-bump.

We would expect that an increase in the C/O mass fraction, at least in the driving region, would lead to the eventual driving of the modes, as it is known that high metallicities help to increase the potential instability of the models.

Models 1 and 8 have the Z-bump playing a role in the destabilization of the {\em p}-modes\footnote{Models marked with a lowercase z in Table~\ref{tab:piz} mean that the model barely reached the 'tendency to instability', considered as such when the growth rate values of the modes reach at least -0.5.}. Model 8 presented an oscillatory profile of the growth rate with frequency, which lead us to analyse it in more detail, revealing mode trapping of {\em g}- and {\em p}-modes \citep{crl09b}.

While doing this study we were aware that standard sdB evolutionary models with solar and uniform metallicities were not able to drive modes, and crude increases in the metallicity first \citep{charpinet96} and the so-called second generation models next \citep{charpinet97b} were needed to excite pulsations and reproduce theoretically the observed instability strip. The latter take diffusion (radiative levitation and gravitational settling) into account to produce a non-uniform profile of iron with depth, which is confirmed to be a key ingredient in driving pulsations \citep{charpinet09}. 

Due to our negative results in exciting modes in this first batch of models, we aimed at building models as rich in metals as possible to test if it was possible to excite modes at some point via a $\kappa$-mechanism associated with a C/O partial ionization zone. Therefore, a last batch of models (models 10 to 16) was built in which an {\em ad hoc} extra-mixing was added to increase the metallicity in the envelope.

All these mixed-models have a tendency to driving of the high-radial order {\em g}-modes (see Table~\ref{tab:piz}), caused by the Z-bump (this was only found previously for model 7), the C/O partial ionization zone, or a combination of both, the C/O being responsible for the tendency to driving of modes with lowest frequencies.

We finally achieved the instability of several models in the evolutionary sequence of model 10, that define a narrow instability band in the range $45\,000 \leq T_\mathrm{eff}\leq 54\,000$~K, $4.2 \leq \log g \leq 4.5$, metallicity $ 0.14 \leq Z \leq 0.15$ and periods around 20~min, 45~min and 1~h. The instability is caused by a classical $\kappa$-mechanism due to the opacity bump of the iron-peak elements, that excite high-radial order {\em g}-modes in the star. Further time-evolved models in the same sequence were found stable. 

Model 15, with $T_\mathrm{eff}=90\,000$~K, $\log g=5.86$ and metallicity Z=0.19, was found to have unstable high-radial order {\em g}-modes due to driving from a combination of a classical $\kappa$-mechanism in the C/O partial ionization zone at $T \sim 2~10^6$~K, a $\kappa$ mechanism operating at $T \sim 2 ~10^7$~K probably caused by argon ionization, and a $\kappa$-mechanism associated with an opacity bump that results from the transition from radiative to conductive opacities. This model has one of the largest total metallicity, helium, carbon and oxygen mass fractions within the whole set of models. We found it drives two overstable modes in the period range 7-8~min. 

Models 10.5, 10.6 and 15, with temperatures between 84\,500 and 91\,000 K, show a tendency to instability of the low-radial order {\em p}-modes caused by the Z-bump. This behaviour is similar to that found in model 1, also a high temperature model ($T_\mathrm{eff} = 79\,000$ K). This is a somewhat unexpected result, as for such high $T_\mathrm{eff}$ the location of the Z-bump is quite high in the envelope of the star, where the driving efficiency is lower due to the lower density. Fortunately, this result is confirmed by actual driving of {\em p}-modes through a $\kappa$ mechanism associated to the Z-bump is achieved by \citet{fontaine08} for two hot sdO models with $T_\mathrm{eff} =$ 71\,000 and 81\,000~K, which include radiative levitation of iron. Those models have high temperatures fitting the first determination of physical parameters of J1600+0748, $T_\mathrm{eff}=71\,000 \pm 2725 $~K and $\log g=5.93 \pm 0.11$, also derived by \citet{fontaine08}.

\section{Discussion and Conclusions}

We carried out a non-adiabatic analysis of sdO equilibrium models to explore their feasibility as pulsators. We analysed 27 sdO models out of 16 different evolutionary sequences and we found the first two sdO evolutionary models ever to drive high-radial order {\em g}-modes. In one model, with $T_\mathrm{eff} =$ 45\,000~K, instability of {\em g}-modes is due to a classical $\kappa$-mechanism in the partial ionization zone of the iron-peak elements at $T \sim 2~10^5$~K. This shows that radiative levitation is not an absolute pre-requisite for pulsations of a model sdO star as long as the metallicity is increased to what are probably unrealistic values. Inclusion of radiative levitation of iron remains necessary for complete asteroseismological analysis of hot subdwarfs. In the second model, with $T_\mathrm{eff} =$ 90\,000~K, {\em g}-modes are found to be unstable due to the combination of driving from three distinct regions in the star. The outermost driving region is located at $T \sim 2~10^6$~K, where the iron-peak opacity bump is strongly augmented by contributions from ionization of carbon and oxygen, which we have have enhanced in abundance by ad-hoc extra-mixing during the evolution after the ZAHB. The second driving region is located at $T \sim 2~10^7$~K where a subtle opacity bump develops due to the ionization of argon. The innermost driving region is located at the transition between radiative and conductive energy transport, and may be of interest as a tool to explore energy transport processes in dense hot regions.

All of our sdO models, except one, present a certain tendency to instability for high-radial order {\em g}-modes, with typical oscillation frequencies between $\sim$0.5 and 2~mHz (i.e. periods $\sim$500--2000~s). For most of the models this tendency to driving in usually due to the occurrence of a driving zone at the location of the C/O partial ionization zone. However, quasi-excitation due to the heavy elements partial ionization zone or a combination of both also occurred.

The fact that we found actual driving due to a $\kappa$-mechanism associated with C/O opacity in only one model may be due to the deep location of the C/O partial ionization zone within the star, where the high ratio of the thermal to dynamical time scales impedes efficient heat interchange. Hence, only hotter models with shallower partial ionization zones, such as model 15 for which we achieved actual driving, may be able to overcome this problem.

The tendency to instability of {\em p}-modes was not so common, and was achieved for only 5 out of 27 models. We did not achieve actual driving of {\em p}-modes, those observed in the only sdO pulsator known to date, in any of our models. In some of the hotter models, we found a tendency to driving of low-radial order {\em p}-modes, always triggered by the $\kappa$-mechanism associated with the Z-bump. We partially attribute the inability to get actual {\em p}-mode driving to the shallow location of the heavy elements partial ionization zone, where the density is low, making its contribution to the work integral not sufficient to drive the modes. However, this difficulty has been now overcome by the results of \citet{fontaine08} who have achieved the actual driving of {\em p}-modes for hot models which include radiative levitation and gravitational settling of iron.

Finally, we would like to remark that, as shown by \citet{fontaine08} and \citet{charpinet09}, inclusion of radiative levitation of iron in the models has revealed essential in achieving the actual excitation of {\em p}-modes. The determination of physical parameters (\citealt{fontaine08}; \citealt{crl09a}) for the unique pulsating sdO star to date provides us with an excellent clue to pursue observational searches for new sdO pulsators, with the aim of characterising their instability strip. Envisaged problems are the lack of catalogued sdOs with high effective temperatures, as historically, atmosphere grids reached only up to about 55\,000~K. The role of recent sky surveys such as the Sloan Digitally Sky Survey\footnote{http://www.sdss.org/} and new tools as those offered by the Virtual Observatory\footnote{http://www.ivoa.net} are crucial for spectroscopic and photometric searches of new sdO candidates. The road ahead is challenging, as candidates will be mostly faint (magnitudes at least over 16 or 17) which will pose difficulties in achieving high S/N light curves that allow to detect the lowest amplitude modes. Strategies as propose sdO candidates to be observed by asteroseismologic satellites such as CoRoT or KEPLER have to be considered.

\section*{Acknowledgments}
CRL acknowledges an {\em \'Angeles Alvari\~ no} contract of the regional government {\em Xunta de Galicia}. This research was also supported by the Spanish Ministry of Science and Technology under project ESP2004--03855--C03--01 and by the \emph{Junta de Andaluc\'ia} and the {\em Direcci\'on General de Investigaci\'on (DGI)} under project AYA2000-1559. AM acknowledges financial support from a {\em Juan de la Cierva} contract of the Spanish Ministry of Education and Science. RO is supported by the Research Council of Leuven University through grant GOA/2003/04.

\end{document}